\documentclass{amsart}	
\usepackage{marvosym}
\usepackage{amsmath,amsfonts,amsthm,mathtools,cases,amssymb}
\usepackage{tikz-cd}
\usepackage{adjustbox}
\usepackage{empheq}
\usepackage{enumerate} 					
\newtheorem{theorem}{Theorem}[section]
\newtheorem{lemma}[theorem]{Lemma}
\newtheorem{proposition}[theorem]{Proposition}
\newtheorem{remark}[theorem]{Remark}
\newtheorem{definition}[theorem]{Definition}

\newtheorem{example}[theorem]{Example}	
\author{Joe Pallister}
\begin{document} 
\title{Period $2$ quivers and their $T$- and $Y$-systems}
\begin{abstract}
We give a general definition for period $2$ quivers before making some prudent assumptions to reduce the number of possibilities. Finding these quivers requires solving a complicated system of equations between the number of arrows in the quiver. We find all solutions of these systems for quivers with $N=3,4,5$ vertices and give some solutions for $N=6$. We then consider the $T$- and $Y$-systems associated with these quivers, some of which exhibit periodic quantities which lead to reductions; in a few cases we obtain the Somos-4 and Somos-5 recurrences.
\end{abstract}
\maketitle
\section{Introduction}
Quiver mutation was defined in \cite{clusteri} and is an operation for transforming one quiver $Q$ into another, provided we do not have any loops or $2$-cycles. The mutation is denoted $\mu_k$ and can be performed at any vertex $k$. It is defined in $3$ steps: 
\begin{enumerate}
\item For each length two path $i\rightarrow k \rightarrow j$ add a new arrow $i\rightarrow j$.
\item Reverse the direction of all arrows entering or exiting $k$.
\item Delete all 2-cycles that have appeared.
\end{enumerate} 
If we write $b_{i,j}$ and $b'_{i,j}$ for the number of arrows from $i$ to $j$ in $Q$ and $\mu_k(Q)$ respectively then this definition is equivalent to
\begin{equation}\label{quivermutation}
b'_{i,j}=
\begin{cases}
-b_{i,j} & i=k \mbox{ or } j=k, \\
b_{i,j}+\tfrac{1}{2}(|b_{i,k}|b_{k,j}+b_{i,k}|b_{k,j}|) & \mbox{ otherwise. }
\end{cases}
\end{equation}
We will often express the information given by the $b_{i,j}$ as a skew-symmetric matrix $B=(b_{i,j})$ called the exchange matrix (or $B$ matrix) for $Q$.

In \cite{fordymarsh} the authors look for quivers which are the same after a single mutation, up to a certain relabelling. Explicitly, for a quiver $Q$ with $N$ vertices, we should have
\begin{equation}\label{period1eqtn}
\rho\mu_1(Q)=Q
\end{equation}
where the permutation $\rho$ relabels the vertex labels of the quiver by 
\begin{equation}\label{introrho}
\rho=(1,2,3,\ldots,N-1,N)
\end{equation}
and it has been assumed, without loss of generality, that the mutation should occur at the vertex labelled ``$1$". These quivers are then called period $1$. Fordy and Marsh were able to give the following classification of period $1$ quivers.
\begin{theorem}\label{fordymarshtheorem}
A quiver is period one if and only if the $b_{i,j}$ satisfy 
\begin{equation}\label{fordymarsh1}
b_{1,j}=b_{1,N+2-j}
\end{equation}
for $j=2,\ldots,N$ and
\begin{equation}\label{fordymarsh2}
b_{i,j}=b_{i-j+1,1}+\sum_{m=2}^j\epsilon_{m,1,i-j+m}
\end{equation}
for all $i>j$. Here 
\begin{equation}
\epsilon_{i,j,l}:=\frac{1}{2}\left(|b_{i,j}|b_{j,l}+b_{i,j}|b_{j,l}|\right)
\end{equation}
is the expression that appears in the second case of (\ref{quivermutation}).
\end{theorem} 
This theorem says that each of the $b_{i,j}$ is determined by the arrows at vertex $1$, by equation (\ref{fordymarsh2}), and that these satisfy the ``palindrome" property (\ref{fordymarsh1}). Conversely any choice of $b_{1,j}$ satisfying (\ref{fordymarsh1}) can be used to generate a period $1$ quiver. Naturally one can ask if this tidy result can be generalised to quivers with larger periods.

Fordy and Marsh give the following definition of period $2$ quivers,
\[
\rho\mu_2\mu_1(Q)=Q
\]
where we mutate at vertices $1$ and $2$, and use the same permutation (\ref{introrho}). For us, this definition seems a bit restrictive, a more general definition might be
\begin{equation}\label{period2permutation}
\sigma\mu_k\mu_j(Q)=Q
\end{equation}
for any permutation $\sigma$ and any pair of vertices $j,k$. In fact, after a relabelling of $Q$, we can assume that the first mutation occurs at $1$ and that the permutation $\sigma$ is given by taking $(1,2,3,\ldots,N)$ and ``splitting" it into as many permutations as we like, without reordering the integers. Two examples are:
\[
(1,2,3)(4,5,6,7)(8,9,\ldots,N), \qquad (1)(2,3)(4,5)(6,7,\ldots,N-1)(N).
\]
As an example with $N=4$, for $\sigma=(1,2)(3,4)$ and $k=4$, the quiver
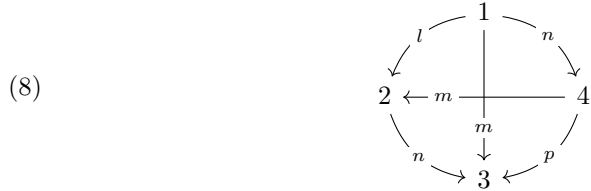
\begin{equation}
\begin{tikzcd}
& 1\arrow[dl, "l" description, bend right]\arrow[dr, "n" description, bend left]\arrow[dd, "m" description, near end]  & \\
2\arrow[dr, "n" description, bend right] & & 4\arrow[ll, "m" description, near end]\arrow[dl, "p" description, bend left]  \\
& 3 & 
\end{tikzcd}
\end{equation}
satisfies $\sigma\mu_4\mu_1(Q)=Q$. 

In addition to quiver mutation, there is also an operation defined in \cite{clusteri} called cluster mutation. We first attach variables $x_i$ to each vertex $i$ of $Q$, each is called an initial cluster variable and the set $\mathbf{x}_0:=\{x_1,\ldots,x_N\}$ is called the initial cluster. For any vertex $k$ the mutation $\mu_k$, in addition to giving a new quiver as defined above, gives a new variable $x'_k:=\mu_k(x_k)$ defined by
\begin{equation}\label{clustermutationformula}
x'_k:=\frac{1}{x_k}\left(\prod_{i\rightarrow k} x_i+\prod_{i\leftarrow k}x_i\right)
\end{equation}
where the products are over the arrows into (and out of) $k$ in $Q$. This gives a new cluster $\mu_k(\mathbf{x}_0):=\{x_1,\ldots,x'_k,\ldots,x_N\}$. We write the exchange matrix for $Q$ as $B_0$ and call the pair $(B_0,\mathbf{x}_0)$ the initial seed. We can consider mutation at $k$ as an operation giving a new seed 
\[
\mu_k(B_0,\mathbf{x}_0)=(\mu_k(B_0),\mu_k(\mathbf{x}_0))
\]
After this we are free to mutate further, as many times as we like. We define the set of cluster variables as the set of elements of the clusters obtained by applying any sequence of mutations to the initial seed. The cluster algebra is the $\mathbb{Z}$-algebra generated by the set of cluster variables.

In general, as we perform further mutations, the form of the right hand side of (\ref{clustermutationformula}) will change. However, for period $1$ quivers, Fordy and Marsh showed that when performing the sequence of mutations $\mu_1,\mu_2,\mu_3,\ldots$ the right hand side will always look the same. By this we mean that if we take initial cluster variables $x_1,\ldots,x_N$ and call the cluster variable obtained after $i$ mutations $x_{N+i}$ then the relations \eqref{clustermutationformula} for the sequence $(x_n)_{n\in\mathbb{Z}}$ reduce to a recurrence relation, as follows.
\begin{theorem}
The recurrences obtained from period $1$ quivers are precisely those of the form 
\begin{equation}\label{period1reccurence}
x_{n+N}x_n=\prod_{i=1}^{N-1}x_{n+i}^{[b_{1,i+1}]}+\prod_{i=1}^{N-1}x_{n+i}^{-[b_{1,i+1}]}
\end{equation}
with $b_{1,j}=b_{1,N+2-j}$ for $j=2,\ldots,N$. Here the square brackets are defined by
\[
[\alpha]=\begin{cases}
\alpha & \quad \alpha\geq 0, \\
0 & \quad \alpha<0.
\end{cases}
\]
\end{theorem} 
These recurrences fit into Nakanishi's definition of $T$-systems for a quiver with any period, \cite{nakanishi}, which he gave to generalise classical $T$-systems. These are obtained by applying a more complicated sequence of mutations to the initial quiver and lead to a system of recurrences, rather than just one. In fact, to obtain these systems we will mutate at each of the vertices in the $\sigma$ orbit of $1$ and $k$. We will not mutate at the vertices outside of these orbits so the cluster variables at those vertices will not change; these cluster variables will essentially be coefficients in the $T$-system. To simplify our search we assume that we mutate at every vertex, in which case the permutation $\sigma$ of \eqref{period2permutation} will be one of two forms:
\[
\sigma=(1,2,\ldots,k,\ldots,N), \qquad \sigma=(1,2,\ldots,k-1)(k,\ldots,N)
\]
and the $T$-systems look quite different for each. In this work we make one final simplification: If $Q$ satisfies the period $2$ equation \eqref{period2permutation} then so does $\mu_1(Q)$ (up to a relabelling). Hence it is enough to find either $Q$ or $\mu_1(Q)$. This allows us to restrict our search to fewer choices of $k$. The final form of the period $2$ equation we take is as follows.
\begin{lemma}
Up to a relabelling of vertices a quiver $Q$ is period $2$ if it, or $\mu_1(Q)$, satisfies 
\begin{equation}\label{period2final}
\sigma\mu_k\mu_1(Q)=Q
\end{equation}
for either
\begin{align*}
&(i) & &\sigma=(1,2,\ldots,N), & & k=2,\ldots,\lfloor\tfrac{N+1}{2}\rfloor, \\
&(ii) & &\sigma=(1,2,\ldots,k-1)(k,\ldots,N), & & k=2,\ldots,\lfloor\tfrac{N+2}{2}\rfloor.
\end{align*}
\end{lemma}
This result is a composite of Lemmas \ref{reorderinglemma} and \ref{pairslemma}. A large portion of this paper is devoted to finding solutions of \eqref{period2final} with these assumptions, but it is not clear how to classify period $2$ quivers in general. Instead we calculate the period $2$ quivers for a low number of vertices, of which there are surprisingly many. We give these results in Theorems \ref{N=3theorem}, \ref{N=4theorem}, \ref{5nodeperiod2theorem}, \ref{feniatheorem} and \ref{N=6theorem}.

A special case of the $T$-systems obtained from period $1$ quivers,  (\ref{period1reccurence}), is the recurrence
\begin{equation}\label{Atyperecurrence}
x_{n+N}x_n=x_{n+p}x_{n+q}+1
\end{equation}
where $N=p+q$. This is obtained from mutating orientations of affine $\tilde{A}_{q,p}$ diagrams, with $q$ arrows pointing clockwise and $p$ anticlockwise. The recurrence (\ref{Atyperecurrence}) was shown in \cite{fordymarsh,fordyhone} to have expressions
\[
J_n:=\frac{x_{n+2q}+x_n}{x_{n+q}}, \qquad \tilde{J}_n:=\frac{x_{n+2p}+x_n}{x_{n+p}}
\]
which are period $p$ and $q$ respectively. These are used to show that there exists linear relations between the iterates of (\ref{period1reccurence}) and that, for even $N$, the ``cluster map" defined by \eqref{Atyperecurrence} is Liouville integrable, with the $J_n$ and $\tilde{J}_n$ used to construct conserved quantities for this map. In \cite{pallisterlinear} this approach was applied to the $T$-systems from affine $\tilde{D}_N$ and $\tilde{E}$ type quivers, to show that they also have periodic quantities that lead to to linear relations between the cluster variables and that reductions of their respective cluster maps are also integrable, with the exception of $\tilde{D}_N$ for odd $N$. The resulting linear relations were also found in \cite{kellerscherotzke}, using links between cluster algebras and representation theory. The existence of such linear relations, for $\tilde{A}$ and $\tilde{D}$ type, was also proved in \cite{frises} using friezes. In this work we show that a handful of the systems from period $2$ quivers also have periodic quantities that, in many cases, allow us to give a simpler expression of the system. In some cases the system reduces to a single recurrence, including the Somos-4 and Somos-5 recurrences studied in \cite{hone2005elliptic,hone2007sigma}. We leave the question of the integrability of these systems for future study.

In addition to $T$-systems, there is a corresponding system, for any periodic quiver, known as a $Y$-system: we attach variables $y_i$ to each vertex and $\mu_k$ gives new $y$ variables defined by
\begin{equation}\label{ymutation}
\mu_k(y_j)=\begin{cases}
y_k^{-1} & j=k, \\
y_j\left(1+y_k^{\mathrm{sgn}(b_{jk})}\right)^{b_{jk}} & j\neq k,
\end{cases}
\end{equation}
where $\mathrm{sgn}$ denotes the sign function. We can include this information in the seeds defined above: the initial seed is $(B_0,\mathbf{x}_0,\mathbf{y}_0)$ and mutation gives a new seed 
\[
\mu_k(B_0,\mathbf{x}_0,\mathbf{y}_0)=(\mu_k(B_0),\mu_k(\mathbf{x}_0),\mu_k(\mathbf{y}_0))
\]
The $Y$-system is the system of recurrences in the $y$ variables obtained from same sequence of mutations we used to obtain the $T$-system. In general, calculating the $T$- and $Y$-systems directly can be painful, but fortunately \cite{nakanishi} gives a compact formula, which we detail later as we apply it to period $2$ quivers.

For period $1$ quivers the $Y$-system is given by
\begin{equation}\label{periodoneYsystem}
y_{n+N}y_n=\frac{
\prod_{j=1}^{N-1}(1+y_{n+j})^{[-b_{1,j+1}]}
}{
\prod_{j=1}^{N-1}(1+y_{n+j}^{-1})^{[b_{1,j+1}]}
}
\end{equation}
The relationship between (\ref{period1reccurence}) and (\ref{periodoneYsystem}) was studied in \cite{honeinoue}. There the authors introduce a new variable $Z_n$ that multiplies the right hand side of \eqref{period1reccurence} and call this new system the $T_Z$ system. They then show that, for a solution of the $T_z$ system, taking
\[
\bar{y}_n=\prod_{j=1}^{N-1}x_{n+j}^{-b_{1,j+1}}
\] 
gives a solution of the $Y$-system \eqref{periodoneYsystem} if and only if the $Z$ variables satisfy 
\[
\prod_{j=1}^{N-1}Z_{n+j}^{-b_{1,j+1}}=1
\]
A similar result as holds for general $T_Z$- and $Y$-systems, the details of which can be found in Remark \ref{TZremark}.

The link between $Y$-systems and cluster algebras was first given in \cite{clustersiv} for bipartite quivers. Connections between cluster algebras and a further type of system, $Q$-systems, were shown in \cite{kedem}. There are also relations with other integrable systems; cluster algebras have been constructed with cluster variables satisfying the discrete Lotka-Volterra and discrete Liouville equations \cite{inouenakanishi}, the Hirota-Miwa equation \cite{okubo2013discrete}, and the discrete mKdV and discrete Toda equations \cite{okubo}. These papers by Okubo and \cite{honeinoue} also demonstrate that certain discrete Painlev\'e equations arise from cluster algebras. A review of many of the other connections between $T$- and $Y$-systems (including their appearance in cluster algebras) and integrable systems can be found in \cite{kuniba2011review}. 

Many authors have also been interested in the periodicities of such systems, inspired by the Zamolodchikov conjecture \cite{zamolodchikov}, which predicts the periodicity of $Y$-systems associated with Dynkin $ADE$ quivers. A proof of this was given in \cite{Ysystemsassociahedra} using the then nascent theory of cluster algebras. The periodicity of the corresponding $T$-systems follows from the ``finite type cluster algebras" classification in \cite{clusterii}: a cluster algebra has finitely many cluster variables if and only if the initial quiver is mutation equivalent to a Dynkin $ADE$ quiver. Studies of generalised forms of Zamolodchikov's conjecture can be found in, for example, \cite{inoue2010periodicities,kellerperiodicity,galashin}. 

This paper is arranged as follows: In Section 2 we discuss our definition of period $2$ quivers and what assumptions we can make about the vertices we mutate at and the permutation we use. We then write down the system of equations satisfied by the $B$ matrix entries for a period $2$ quiver. Afterwards we define and obtain the $T$- and $Y$-systems related to these quivers. In Sections 3, 4, 5 and 6 we find all solutions of the period $2$ equation for quivers with $N=3,4,5$. In Section 7 we find some solutions for $N=6$. Finally in Section 8 we discuss some of the $T$- and $Y$-systems for the quivers we have found, which can be simplified due to the presence of periodic quantities. 
\section{Period $2$ quivers}
We begin here by justifying the assumptions we make about the period $2$ equation \eqref{period2permutation}, which will allow us to greatly reduce the number of cases we need to check. In Section \ref{systemsection} we write down the period $2$ equation in terms of the $B$ matrix entries $b_{i,j}$, which is the system of equations we need to solve to find period $2$ quivers. Since we are interested in the $T$- and $Y$-systems that can be obtained from these, in Section \ref{Naksection} we give Nakanishi's construction of $T$- and $Y$-systems for quivers with any period. In Section \ref{firstTsection} we describe how the $T$-system looks for period $2$ quivers with the permutations $(1,2,\ldots,N)$ and $(1,\ldots,k-1)(k,\ldots,N)$. Finally in \ref{Ysystemssubsection} we write down the corresponding $Y$-systems. 
\subsection{Assumptions about the permutation $\sigma$}\label{assumptionssection}
Before starting the search for period $2$ quivers, we first justify our assumptions about the defining equation $\sigma\mu_k\mu_1(Q)=Q$.
\begin{lemma}\label{reorderinglemma}
Without loss of generality we can assume that the first mutation occurs at vertex $1$ and that the permutation $\sigma$ is given by 
\begin{equation}\label{sigmaform}
\sigma=\prod_{i=1}^{\alpha}(\sigma_{i_1}\sigma_{i_2}\ldots\sigma_{i_{\beta_i}})
\end{equation}
where $\alpha$ and $\beta_i$ are integers and the $\sigma_{i_j}$ satisfy
\begin{equation}\label{sigmaconditions}
\sigma_{i_{j+1}}=\sigma_{i_j}+1, \qquad \sigma_{{(i+1)}_1}=\sigma_{i_{\beta_i}}+1, \qquad \sigma_{1_1}=1.
\end{equation}
Furthermore, if the second mutation $k$ appears in a different cycle to $1$ then we can assume that it appears at the start of the second cycle, i.e. in (\ref{sigmaform}) we have $\sigma_{2_1}=k$. Equivalently $\sigma$ is given by
\begin{equation}\label{sigmaformdifferent}
\sigma=(1,2,\ldots,k-1)(k,k+1,\ldots,\sigma_{2_{\beta_2}})\prod_{i=3}^{\alpha}(\sigma_{i_1}\sigma_{i_2}\ldots\sigma_{i_{\beta_i}})
\end{equation}
satisfying (\ref{sigmaconditions}).
\begin{proof}
Let the general permutation in (\ref{period2permutation}) be given by
\[
\prod_{i=1}^{\alpha}(\tilde{\sigma}_{i_1}\tilde{\sigma}_{i_2}\ldots\tilde{\sigma}_{i_{\beta_i}}).
\]
We let $\sigma'$ be the permutation given by the simultaneously relabelling $\tilde{\sigma}_i\mapsto i$, for each $i$. This satisfies (\ref{sigmaconditions}). Let $j'$ be the first vertex we mutate under this new labelling. We now subtract $j'-1$ from every vertex label to ensure that the first mutated vertex is $1$, without violating conditions (\ref{sigmaconditions}). 

Finally if $k$ appears in a different cycle to $1$ then we can apply this lemma to the permutation that occurs after the first cycle in (\ref{sigmaform}):
\begin{equation}
\prod_{i=2}^{\alpha}(\sigma_{i_1}\sigma_{i_2}\ldots\sigma_{i_{\beta_i}})
\end{equation}
to bring $k$ to the start, without affecting the order.
\end{proof}
\end{lemma}
Because of this we can instead look at the equation 
\begin{equation}\label{period2eqtn}
\sigma\mu_k\mu_1(Q)=Q
\end{equation}
where $\sigma$ is of the form (\ref{sigmaform}) or (\ref{sigmaformdifferent}).
\begin{example}
For $N=5$ the possible permutations are given by the ways we can take
\[
(1,2,3,4,5)
\]
and separate it into cycles by adding brackets, without reordering the numbers. We have the following options
\begin{align*}
&(1,2,3,4,5) & &(1,2,3,4)(5) & &(1,2,3)(4,5) & &(1,2)(3,4,5) \\
&(1)(2,3,4,5) & &(1,2,3)(4)(5) & &(1,2)(3,4)(5) & &(1)(2,3,4)(5) \\
&(1,2)(3)(4,5) & &(1)(2,3)(4,5) & &(1)(2)(3,4,5) & &(1,2)(3)(4)(5)\\
&(1)(2,3)(4)(5) & &(1)(2)(3,4)(5) & &(1)(2)(3)(4,5) & &(1)(2)(3)(4)(5)
\end{align*}
There are $4$ choices of $k$ for $(1,2,3,4,5)$ and one for the rest of the permutations (by taking $k$ as the first entry in the second cycle).
\end{example}
\begin{example}
The quiver, given in \cite{fordymarsh} equation (11),
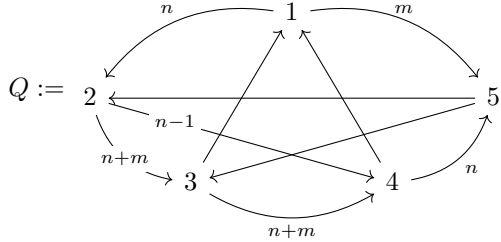
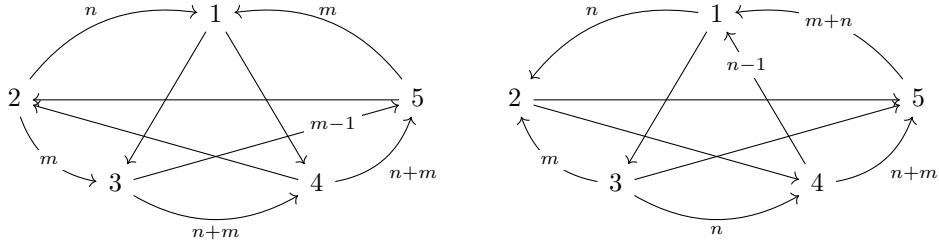
\[
Q:=
\begin{tikzcd}
& & 1  \arrow[drr, "m" above, bend left] \arrow[dll, "n" above left, bend right]& & \\
2 \arrow[dr, "n+m" description, bend right] \arrow[drrr, "n-1" description, near start]& & & & 5\arrow[dlll] \arrow[llll] \\
& 3\arrow[rr, "n+m" below, bend right]\arrow[uur]  & & 4  \arrow[ur, "n" below right, bend right]\arrow[uul] &
\end{tikzcd}
\]
is mutated to 
\[
\begin{tikzcd}
& & 1 \arrow[ddl] \arrow[ddr] & & \\
2 \arrow[urr, "n" above left, bend left]\arrow[dr, "m" description, bend right] & & & & 5 \arrow[ull, "m" above, bend right]\arrow[llll] \\
& 3\arrow[rr, "n+m" below, bend right] \arrow[urrr, "m-1" description, near end] & & 4 \arrow[ulll] \arrow[ur, "n+m" below right, bend right] &
\end{tikzcd}
\qquad 
\begin{tikzcd}
& & 1 \arrow[ddl]  \arrow[dll, "n" above left, bend right]& & \\
2 \arrow[rrrr] \arrow[drrr]& & & & 5 \arrow[ull, "m+n" description, bend right] \\
& 3\arrow[rr, "n" below, bend right] \arrow[urrr] \arrow[ul, "m" description, bend left]& & 4\arrow[uul, "n-1" description, near end]  \arrow[ur, "n+m" below right, bend right] &
\end{tikzcd}
\]
by $\mu_1$ and then $\mu_2$. We see that the permutation $\sigma=(1,4,2,5,3)$ applied to $\mu_2\mu_1(Q)$ gives $Q$ back. In this case the relabelling $\sigma'$ is given by 
\[
1\mapsto 1, \qquad 4\mapsto 2, \qquad 2\mapsto 3, \qquad 5\mapsto 4, \qquad 3\mapsto 5
\]
so
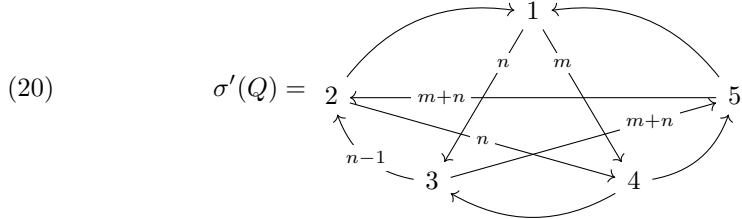
\begin{equation}\label{firstexamplequiver}
\sigma'(Q)=
\begin{tikzcd}
& & 1 \arrow[ddl, "n" description, near start] \arrow[ddr, "m" description, near start] & & \\
2 \arrow[urr, bend left] \arrow[drrr, "n" description]& & & & 5\arrow[ull, bend right] \arrow[llll, "m+n" description, near end] \\
& 3 \arrow[urrr, "m+n" description, near end] \arrow[ul, "n-1" description, bend left]& & 4 \arrow[ll, bend left] \arrow[ur, bend right] &
\end{tikzcd}
\end{equation}
has the property that $\rho\mu_3\mu_1\sigma'(Q)=\sigma'(Q)$, so that $\sigma'(Q)$ satisfies (\ref{period2eqtn}).
\end{example}
As mentioned in the introduction, when we obtain the $T$- and $Y$-systems from these quiver we will mutate at every vertex in the $\sigma$ orbit of $1$ and $k$, so vertices outside of these orbits will be frozen. In our work we will assume that there are no frozen vertices, so we can update the period $2$ equation to
\begin{equation}\label{period2eqtnfinal}
\sigma\mu_k\mu_1(Q)=Q
\end{equation}
where 
\[
\sigma=(1,2,\ldots,N) \qquad \mathrm{or} \qquad \sigma=(1,\ldots,k-1)(k,\ldots,N).
\]
We give more details about the construction of $T$- and $Y$-systems in Section \ref{Naksection}. 

Finally we note that, since $Q$ is period $2$, $\mu_1(Q)$ is also period $2$. This allows us to further reduce our search.
\begin{lemma}\label{pairslemma} 
If $Q$ satisfies \eqref{period2eqtnfinal} then we have the following two cases.
\begin{enumerate}[(i)]
\item For $\sigma=(1,\ldots,N)$, after a relabelling, the quiver $Q':=\mu_1(Q)$ satisfies
\[
\sigma\mu_{N-k+1}\mu_1(Q')=Q'.
\]
\item For $\sigma=(1,\ldots,k-1)(k,\ldots,N)$, after a relabelling, $Q'$ satisfies
\[
\sigma'\mu_{N-k+2}\mu_1(Q')=Q'
\]
where $\sigma'=(1,2,\ldots,N-k+1)(N-k+2,\ldots,N)$.
\end{enumerate}
\begin{proof}
We take $Q':=\mu_1(Q)$. By \eqref{period2eqtnfinal}, we have
\[
Q'=\mu_1\sigma\mu_k(Q')=\sigma\mu_{\sigma^{-1}(1)}\mu_k(Q').
\]
Now we define $Q''$ by $Q'=\sigma''(Q'')$ where $\sigma''$ is the permutation $\sigma''(i)=i+k-1$ (with labels taken mod $N$), then we have
\begin{equation}\label{reduction}
Q''=(\sigma'')^{-1}\sigma\mu_{\sigma^{-1}(1)}\mu_k\sigma''(Q'')=(\sigma'')^{-1}\sigma\sigma''\mu_{(\sigma'')^{-1}\sigma^{-1}(1)}\mu_{(\sigma'')^{-1}(k)}(Q'').
\end{equation}
In the case with $\sigma=(1,2,\ldots,N)$ we have 
\[
\sigma=(\sigma'')^{-1}\sigma\sigma'',\qquad(\sigma'')^{-1}(k)=1, \qquad (\sigma'')^{-1}\sigma^{-1}(1)=N-k+1
\]
so here \eqref{reduction} is
\[
\sigma\mu_{N-k+1}\mu_1(Q'')=Q''
\]
so $Q'$ satisfies the required equation, up to the relabelling $\sigma''$. This is part (i) of the lemma. 

For $\sigma=(1,\ldots,k-1)(k,\ldots,N)$, however, we have $(\sigma'')^{-1}\sigma\sigma''=\sigma'$ where
\begin{equation}\label{strangeperm}
\sigma':=(1,2,\ldots,N-k+1)(N-k+2,\ldots,N)
\end{equation}
and $(\sigma'')^{-1}\sigma^{-1}(1)=N$, so \eqref{reduction} is
\begin{equation}\label{how}
\sigma'\mu_{N}\mu_1(Q'')=Q''.
\end{equation}
As discussed in Lemma \ref{reorderinglemma}, we can assume that the second mutation occurs at $N-k+2$ by relabelling the vertices of the second cycle in \eqref{strangeperm}, while fixing $\sigma'$. Hence $Q'$, up to these two relabellings, satisfies part (ii).
\end{proof}
\end{lemma}
Since both $Q$ and $\mu_1(Q)$ are period $2$, it is enough to just find one of them. In case $\sigma=(1,2,\ldots,N)$ we can assume, due to Lemma \ref{pairslemma}, that $k\leq N-k+1$. In case $\sigma=(1,2,\ldots,k-1)(k,\ldots,N)$ we assume that $k\leq N-k+2$. Furthermore taking $k=1$ gives the equation $\sigma(Q)=Q$, which doesn't involve mutation at all, so we ignore it in this paper. We compile all of the reductions of this section as follows.
\begin{lemma}\label{finalformlemma}
To find all period two quivers it is enough to find those satisfying 
\[
\sigma\mu_k\mu_1(Q)=Q
\]
with either
\begin{align*}
&(i) & &\sigma=(1,2,\ldots,N), & & k=2,\ldots,\lfloor\tfrac{N+1}{2}\rfloor, \\
&(ii) & &\sigma=(1,2,\ldots,k-1)(k,\ldots,N), & & k=2,\ldots,\lfloor\tfrac{N+2}{2}\rfloor.
\end{align*}
\end{lemma}
Now that we have the final form of the period $2$ equation we can turn our attention to trying to find solutions. Firstly we need to see what the implications are for the number of arrows in $Q$, the $b_{i,j}$.
\subsection{The defining equations for period $2$ quivers in terms of the $b_{i,j}$}\label{systemsection}
By defining $\tilde{Q}=\mu_k\mu_1(Q)$ the condition (\ref{period2eqtnfinal}) becomes $\sigma\tilde{Q}=Q$, i.e. 
\[
\tilde{b}_{i,j}=b_{\sigma(i),\sigma(j)}
\]
where $b_{i,j}$ and $\tilde{b}_{i,j}$ denote the number of arrows from $i$ to $j$ in $Q$ and $\tilde{Q}$ respectively. Rather than trying to solve $\mu_k\mu_1(Q)=\sigma{Q}$ it seems to be simpler to equate $\mu_1(Q)$ with $\mu_k(\sigma^{-1}{Q})=\mu_k(\tilde{Q})$. We take the definition
\begin{equation}\label{epsilondef}
\epsilon_{i,j,l}:=\frac{1}{2}\left(|b_{i,j}|b_{j,l}+b_{i,j}|b_{j,l}|\right)
\end{equation} 
to tidy the quiver mutation formula (\ref{quivermutation}), which here gives
\[
\mu_1(b_{ij})=
\begin{cases}
-b_{ij} & i \:\mathrm{ or } \:j=1, \\
b_{ij}+\epsilon_{i,1,j} & \mathrm{otherwise}
\end{cases}
\]
and
\[
\mu_k(\tilde{b}_{ij})=
\begin{cases}
-\tilde{b}_{ij} & i \:\mathrm{ or } \:j=k, \\
\tilde{b}_{ij}+\tilde{\epsilon}_{i,k,j} & \mathrm{otherwise.}
\end{cases}
\]
The second of these is equivalent to
\[
\mu_k(\tilde{b}_{ij})=
\begin{cases}
-b_{\sigma(i),\sigma(j)} & i \:\mathrm{ or } \:j=k, \\
b_{\sigma(i),\sigma(j)}+\epsilon_{\sigma(i),\sigma(k),\sigma(j)}& \mathrm{otherwise.}
\end{cases}
\]
The condition $\mu_1(b_{i,j})=\mu_k(\tilde{b}_{i,j})$ then gives the following three cases.
\begin{empheq}[box=\fbox]{align}
-b_{i,j}&=b_{\sigma(i),\sigma(j)}+\epsilon_{\sigma(i),\sigma(k),\sigma(j)} & i=1 \:\mathrm{or}\: j=1 \:\mathrm{and}\: i,j\neq k,\quad\nonumber\\
\quad b_{i,j}+\epsilon_{i,1,j}&=-b_{\sigma(i),\sigma(j)} & i=k \:\mathrm{or}\: j=k \:\mathrm{and}\: i,j\neq 1,\quad\label{brelations}\\
b_{i,j}+\epsilon_{i,1,j}&=b_{\sigma(i),\sigma(j)}+\epsilon_{\sigma(i),\sigma(k),\sigma(j)} & \mathrm{otherwise.}\quad \nonumber
\end{empheq}
These equations appear to be very difficult to solve in general. With some work, however, they can be solved for low $N$, giving many interesting solutions. Once we have found these period $2$ quivers we can consider the associated $T$- and $Y$-systems. We now discuss how these are obtained.
\subsection{$T$- and $Y$-systems for periodic quivers}\label{Naksection}
Here we give the details of Nakanishi's general construction of $T$- and $Y$-systems from periodic quivers before restricting to the period $2$ case. 
\begin{definition}
For a quiver $Q$, we define $\mathbf{i}=(i_1,i_2,\ldots i_r)$ as a sequence of vertices. The mutation $\mu_{\mathbf{i}}$ is the composite of the mutations at each of these vertices, i.e.
\[
\mu_{\mathbf{i}}:=\mu_{i_r}\ldots\mu_{i_2}\mu_{i_1}.
\]  
We let $\nu$ be a permutation acting on the vertex labels of $Q$. We say that the sequence $\mathbf{i}$ is a $\nu$-period of $Q$ if 
\[
b'_{\nu(i),\nu(j)}=b_{i,j}
\]
where $b_{i,j}$ and $b'_{i,j}$ denote the number of arrows between vertices $i$ and $j$ in $Q$ and $\mu_{\mathbf{i}}(Q)$, respectively. We decompose $\mathbf{i}$ into $t$ parts:
\begin{equation}\label{slice}
\mathbf{i}=\mathbf{i}(0)|\mathbf{i}(1)|\ldots|\mathbf{i}(t-1)
\end{equation}
such that, within each part, each of the mutations commute. There can be many ways of doing this, and each is known as a slice of $\mathbf{i}$. 
\end{definition} 
We note that, in Nakanishi's definition, $Q$ satisfies $\mu_{\mathbf{i}}(Q)=\nu(Q)$, so in our notation $\nu=\sigma^{-1}$. We can now apply $\mu_{\mathbf{\nu(i)}}$ to $\mu_{\mathbf{i}}(Q)$ to see that 
\[
\mu_{\mathbf{\nu(i)}}\mu_{\mathbf{i}}(Q)=\nu^2(Q).
\]
Further applications will give
\[
\mu_{\mathbf{\nu^u(i)}}\ldots\mu_{\mathbf{\nu(i)}}\mu_{\mathbf{i}}(Q)=\nu^u(Q)
\]
for all $u\in\mathbb{Z}$.
\begin{definition}
For any integer $u$ we write $u=rt+l$, where $r\in\mathbb{Z}$ and $0\leq l<t$, where $t$ is the number of parts of the slice (\ref{slice}). For $l=0$ we then define the quivers $Q(u)=Q(rt)$ as
\[
Q(rt)=\mu_{\nu^{r-1}(\mathbf{i})}\ldots\mu_{\mathbf{\nu(i)}}\mu_{\mathbf{i}}(Q)
\]
and for $0<l<t$ we define
\[
Q(rt+l)=\mu_{\nu^{r}(\mathbf{i}(l-1))}\ldots\mu_{\nu^{r}(\mathbf{i}(1))}\mu_{\nu^{r}(\mathbf{i}(0))}Q(rt).
\]
We let $b_{ji}(u)$ be the number of arrows from $j$ to $i$ in $Q(u)$. The set of forward mutation points $P_+$ is given by
\[
P_+=\{(i,u)=(i,rt+l)\:\mid\: i\in\nu^r(\mathbf{i}(l))\}.
\]
Finally for each $(i,u)\in P_+$ we let $0<\lambda_{\pm}(i,u)$ be the smallest integers such that 
\[
(i,u\pm\lambda_{\pm})\in P_+.
\]
\end{definition}
To obtain $Q(u+1)$ from $Q(u)$ we mutate at each of the vertices $i$ such that $(i,u)\in P_+$. We denote the new cluster variables obtained by this as $x_i(u+1)$. For a fixed $(i,u)\in P_+$ the point $(i,u-\lambda_{-})$ is the previous point where we have mutated at vertex $i$. Similarly $(i,u+\lambda_{+})$ is the next point. For this work we assume that there are no frozen vertices; we mutate at every vertex at least once.  

With all this we are now able to define the $T$-system, which is given by
\begin{equation}\label{Tsystem}
x_i(u)x_i(u+\lambda_{+}(i,u))=\prod_{(j,v)\in P_+}x_j(v)^{H_+(j,v;i,u)}+\prod_{(j,v)\in P_+}x_j(v)^{H_-(j,v;i,u)}
\end{equation}
where 
\begin{equation}\label{Hpm}
H_{\pm}(j,v;i,u)=
\begin{cases}
\pm b_{ji}(u) & u\in (v-\lambda_{-}(j,v),v), \qquad b_{ji}(u)\gtrless 0, \\
0 & \mathrm{otherwise.}
\end{cases}
\end{equation}
The $Y$-system is defined similarly, as follows:
\begin{equation}\label{Ysystem}
y_i(u)y_i(u+\lambda_+(i,u))=\frac{
\prod_{(j,v)\in P_+} (1+y_j(v))^{G_+(j,v;i,u)}
}{
\prod_{(j,v)\in P_+} (1+y_j(v)^{-1})^{G_-(j,v;i,u)}
}
\end{equation}
where
\[
G_+(j,v;i,u)=\begin{cases}
\mp b_{ji}(v) & v\in(u,u+\lambda_+(i,u)), \qquad b_{ji}\lessgtr 0, \\
0 & \mathrm{otherwise.}
\end{cases}
\]
\begin{remark}\label{TZremark}
The $T_Z$ system of \cite{honeinoue} is given by multiplying the right hand sides of \eqref{Tsystem} by a new variable $Z_i(u)$. Then a solution of the $T_Z$-system gives a solution to the $Y$-system via
\[
\bar{y}_n:=\prod_{(j,v)\in P_+}\frac{x_j(v)^{H_+(j,v;i,u)}}{x_j(v)^{H_-(j,v;i,u)}}
\]
if and only if 
\[
\prod_{(j,v)\in P_+}\frac{Z_j(v)^{H_+(j,v;i,u)}}{Z_j(v)^{H_-(j,v;i,u)}}=1
\]
\end{remark}
We now aim to apply this to our period $2$ quivers. For our defining equation \eqref{period2eqtnfinal} we have $\mathbf{i}=(1,k)$ and $\nu=\sigma^{-1}$. We decompose $\mathbf{i}$ into $t=2$ parts: $\mathbf{i}(0)=1$ and $\mathbf{i}(1)=k$. We have also assumed that the permutation (\ref{sigmaform}) is a product of one or two cycles:
\begin{empheq}{align}
(\mathrm{i}) & \qquad (1,2,\ldots,N), \label{1cycle}\\
(\mathrm{ii}) & \qquad (1,2,\ldots,k-1)(k,k+1,\ldots,k,\ldots,N). \label{2cycles}
\end{empheq}
In each case the $T$ and $Y$-systems (\ref{Tsystem}) take a different form. We describe these in the next $2$ sections.
\begin{remark}
The equations (\ref{brelations}) have been vastly generalised in \cite{mizuno}, Equation 3.9, for quivers with any period. There Mizuno gives a more general definition of $T$-systems and classifies precisely which (generalised) $T$-systems can be obtain from mutation loops. 
\end{remark}
\subsection{$T$-systems for period $2$ quivers}\label{firstTsection}
Here we can use Nakanishi's formula to write down the $T$-systems for period $2$ quivers for the $2$ permutations \eqref{1cycle} and \eqref{2cycles}.
\begin{lemma}\label{firstTsystem}
The $T$-system for the permutation $\sigma=(1,\ldots,N)$ is given by
\begin{align}
z(q)y(q+k-1)
=&\smashoperator[l]{\prod_{p\in(0,N-k+1)}}z(q+p)^{[b_{\overline{1-p},1}(0)]}
&&\smashoperator[l]{\prod_{p\in(-1,k-1)}}y(q+p)^{[b_{\overline{k-p},1}(0)]} \nonumber\\[1em]
+&\smashoperator[l]{\prod_{p\in(0,N-k+1)}}z(q+p)^{[-b_{\overline{1-p},1}(0)]}
&&\smashoperator[l]{\prod_{p\in(-1,k-1)}}y(q+p)^{[-b_{\overline{k-p},1}(0)]} \label{before} \\[1em]
y(q)z(q+N-k+1)
=&\smashoperator[l]{\prod_{p\in(0,N-k+1)}}z(q+p)^{[b_{\overline{1-p},k}(1)]}
&&\smashoperator[l]{\prod_{p\in(0,k)}}y(q+p)^{[b_{\overline{k-p},k}(1)]} \nonumber\\[1em]
+&\smashoperator[l]{\prod_{p\in(0,N-k+1)}}z(q+p)^{[-b_{\overline{1-p},k}(1)]}
&&\smashoperator[l]{\prod_{p\in(0,k)}}y(q+p)^{[-b_{\overline{k-p},k}(1)]} \label{after}
\end{align}
where the bar denotes reduction mod $N$ and $z(q)=x_{\overline{1-q}}(2q)$ and $y(q)=x_{\overline{k-q}}(2q+1)$. The $b_{i,j}(0)$ and $b_{i,j}(1)$ denote the number of arrows in $Q$ and $\mu_1(Q)$ respectively.
\begin{proof}
Since $\mathbf{i}=(1,k)$ and $\mathbf{i}(0)=1$, $\mathbf{i}(1)=k$, we have
\begin{equation}\label{cases}
\nu^r(\mathbf{i}(l))=
\begin{cases}
\overline{1-r} & \quad l=0, \\
\overline{k-r} & \quad l=1
\end{cases}
\end{equation}
where the bar denotes reduction mod $N$. To find the set of forward mutation points $(u,i)$ we write $u=tr+l=2r+l$, where $l=0,1$. For $l=0$ we have $u=2r$ and $i=\overline{1-r}$ due to (\ref{cases}). Similarly for $l=1$ we have $(u,i)=(\overline{k-r},2r+1)$. Hence $P_+$ is given by
\[
P_+=\{(\overline{1-r},2r),\: (\overline{k-r},2r+1)\:\mid\: r\in\mathbb{Z}\}.
\]
We also have
\begin{align}
&\lambda_{-}(\overline{1-r},2r)=2N-2k+1, & &\lambda_{-}(\overline{k-r},2r+1)=2k-1, \nonumber \\
&\lambda_{+}(\overline{1-r},2r)=2k-1, & &\lambda_{+}(\overline{k-r},2r+1)=2N-2k+1. \nonumber
\end{align}
Since there are $2$ types of points in $P_+$, there are $4$ different cases for the $H_{\pm}$ of (\ref{Hpm}) that occur. For example $H_{\pm}(\overline{1-p},2p;\overline{1-q},2q)$ is given by
\begin{align}
\phantom{=}&
\begin{cases}
\pm b_{\overline{1-p},\overline{1-q}}(2q) & 2q\in (2p-(2N-2k+1),2p), \quad b_{\overline{1-p},\overline{1-q}}(2q)\gtrless 0, \\
0 & \mathrm{otherwise,}
\end{cases} \nonumber \\
=&
\begin{cases}
\pm b_{\overline{1-p+q},1}(0) & p\in (q,q+N-k+1), \quad b_{\overline{1-p+q},1}(0)\gtrless 0, \\
0 & \mathrm{otherwise.} 
\end{cases}\nonumber
\end{align}
Here we have used that $b_{\nu^r(i),\nu^r(j)}(2r)=b_{i,j}(0)$ which follows from the periodicity of our quivers. The other $3$ cases for $H_{\pm}$ can be expressed similarly. These appear inside the products in (\ref{Tsystem}), which we can simplify. There are two different cases for the $T$-system, depending on which type of point of $P_+$ appears on the left hand side of \eqref{Tsystem}. For $(i,u)=(\overline{1-q},2q)$ the products on the right hand side 
\[
\smashoperator[l]{\prod_{(j,v)\in P_{+}}}x_j(v)^{H_{\pm}(j,v;\overline{1-q},2q)}
\]
are given by
\begin{align}
&\smashoperator[l]{\prod_{p\in\mathbb{Z}}}x_{\overline{1-p}}(2p)^{H_{\pm}(\overline{1-p},2p;\overline{1-q},2q)}
&&\smashoperator[l]{\prod_{p\in\mathbb{Z}}}x_{\overline{k-p}}(2p+1)^{H_{\pm}(\overline{k-p},2p+1;\overline{1-q},2q)} \nonumber\\[1em]
=&\smashoperator[l]{\prod_{p\in(q,q+N-k+1)}}x_{\overline{1-p}}(2p)^{[\pm b_{\overline{1-p+q},1}(0)]}
&&\smashoperator[l]{\prod_{p\in(q-1,q+k-1)}}x_{\overline{k-p}}(2p+1)^{[\pm b_{\overline{k-p+q},1}(0)]} \nonumber\\[1em]
=&\smashoperator[l]{\prod_{p\in(0,N-k+1)}}x_{\overline{1-p-q}}(2(p+q))^{[\pm b_{\overline{1-p},1}(0)]}
&&\smashoperator[l]{\prod_{p\in(-1,k-1)}}x_{\overline{k-p-q}}(2(p+q)+1)^{[\pm b_{\overline{k-p},1}(0)]} \nonumber
\end{align}
and there is a similar expression if $(i,u)=(\overline{k-q},2q+1)$ appears on the left hand side. We substitute these into the $T$-system (\ref{Tsystem}) and take $z(q)=x_{\overline{1-q}}(2q)$ and $y(q)=x_{\overline{k-q}}(2q+1)$ to prove the lemma.
\end{proof}
\end{lemma}
\begin{lemma}\label{secondTsystemlemma}
The $T$-system for the permutation $\sigma=(1,\ldots,k-1)(k,\ldots,N)$ is given by
\begin{align}
z(q)z(q+k-1)
=&\smashoperator[l]{\prod_{p\in(0,k-1)}}z(p+q)^{[b_{1+p,1}(0)]}
&&\smashoperator[l]{\prod_{p\in(-1,N-k+1)}}y(p+q)^{[b_{k+p,1}(0)]} \nonumber \\[1em]
+&\smashoperator[l]{\prod_{p\in(0,k-1)}}z(p+q)^{[-b_{1+p,1}(0)]}
&&\smashoperator[l]{\prod_{p\in(-1,N-k+1)}}y(p+q)^{[-b_{k+p,1}(0)]} \label{once} \\[1em]
y(q)y(q+N-k+1)
=&\smashoperator[l]{\prod_{p\in(0,k)}}z(p+q)^{[b_{\nu^p(1),k}(1)]}
&&\smashoperator[l]{\prod_{p\in(0,N-k+1)}}y(p+q)^{[b_{k+p,k}(1)]} \nonumber\\[1em]
+&\smashoperator[l]{\prod_{p\in(0,k)}}z(p+q)^{[-b_{\nu^p(1),k}(1)]}
&&\smashoperator[l]{\prod_{p\in(0,N-k+1)}}y(p+q)^{[-b_{k+p,k}(1)]} \label{twice}
\end{align}
where, in the products involving the $z$ variables we have 
\[
\nu^p(1)=
\begin{cases}
1+p & p=1,\ldots,k-2, \\
1 & p=k-1
\end{cases}
\]
and we have taken
\[
z(q)=x_{\nu^q(1)}(2q), \qquad y(q)=x_{\nu^q(k)}(2q+1).
\]
\begin{proof}
In this case the set of forward mutation points $P_+$ is
\[
P_+:=\{(\nu^r(1),2r),\:(\nu^r(k),2r+1)\mid r\in\mathbb{Z}\}
\]
and $\lambda_{\pm}$ is given by
\[
\lambda_{\pm}(j,v)=
\begin{cases}
2(k-1) & j\in(1,\ldots,k-1), \\
2(N-k+1) & j\in(k,\ldots,N).
\end{cases}
\]
The rest of the proof is similar to the previous case.
\end{proof}
\end{lemma}
\subsection{$Y$-systems for period $2$ quivers}\label{Ysystemssubsection}
Here we write down the $Y$-systems for period $2$ quivers. The details for obtaining these from equation (\ref{Ysystem}) are similar to those for the $T$-systems, so we omit them. 
\begin{lemma}
\begin{enumerate}[(i)]
\item
For the permutation $\sigma=(1,2,\ldots,N)$ we have
\begin{align}
&A(q)B(q+k-1)= \nonumber 
\\
&\frac{
\left(\prod_{p\in(0,k)}(1+A(q+p))^{[-b_{1,\overline{1+p}}(0)]}\right)
\left(\prod_{p\in(-1,k-1)}(1+B(q+p))^{[-b_{k,\overline{1+p}}(1)]}\right)
}
{
\left(\prod_{p\in(0,k)}(1+A(q+p)^{-1})^{[b_{1,\overline{1+p}}(0)]}\right)
\left(\prod_{p\in(-1,k-1)}(1+B(q+p)^{-1})^{[b_{k,\overline{1+p}}(1)]}\right)
} \nonumber
\end{align}
and
\begin{align}
&B(q)A(q+N-k+1)= \nonumber
\\
&\frac{
\left(\prod_{p\in(0,N-k+1)}(1+A(q+p))^{[-b_{1,\overline{k+p}}(0)]}\right)
\left(\prod_{p\in(0,N-k+1)}(1+B(q+p))^{[-b_{k,\overline{k+p}}(1)]}\right)
}
{
\left(\prod_{p\in(0,N-k+1)}(1+A(q+p)^{-1})^{[b_{1,\overline{k+p}}(0)]}\right)
\left(\prod_{p\in(0,N-k+1)}(1+B(q+p)^{-1})^{[b_{k,\overline{k+p}}(1)]}\right)
} \nonumber
\end{align}
where we have written $A(q):=Y_{\overline{1-q}}(2q)$ and $B(q):=Y_{\overline{k-q}}(2q+1)$ and the bar denotes reduction mod $N$.
\item
For the permutation $\sigma=(1,\ldots,k-1)(k,\ldots,N)$ we have
\begin{align}
&A(q)A(q+k-1)= \nonumber
\\
&\frac{
\left(\prod_{p\in(0,k-1)}(1+A(q+p))^{[-b_{1,\nu^{-p}(1)}(0)]}\right)
\left(\prod_{p\in(-1,k-1)}(1+B(q+p))^{[-b_{k,\nu^{-p}(1)}(1)]}\right)
}{
\left(\prod_{p\in(0,k-1)}(1+A(q+p)^{-1})^{[b_{1,\nu^{-p}(1)}(0)]}\right)
\left(\prod_{p\in(-1,k-1)}(1+B(q+p)^{-1})^{[b_{k,\nu^{-p}(1)}(1)]}\right)
} \nonumber
\end{align}
and
\begin{align}
&B(q)B(q+N-k+1)= \nonumber
\\
&\frac{
\left(\prod_{p\in(0,N-k+2)}(1+A(q+p))^{[-b_{1,\nu^{-p}(k)}(0)]}\right)
\left(\prod_{p\in(0,N-k+1)}(1+B(q+p))^{[-b_{k,\nu^{-p}(k)}(1)]}\right)
}{
\left(\prod_{p\in(0,N-k+2)}(1+A(q+p)^{-1})^{[b_{1,\nu^{-p}(k)}(0)]}\right)
\left(\prod_{p\in(0,N-k+1)}(1+B(q+p)^{-1})^{[b_{k,\nu^{-p}(k)}(1)]}\right)
} \nonumber
\end{align}
where $A(q):=Y_{\nu^q(1)}(2q)$ and $B(q):=Y_{\nu^q(k)}(2q+1)$. 
\end{enumerate}
\end{lemma} 
\section{Period $2$ quivers with $N=3$ vertices}
We are now ready to start the search for period $2$ quivers. In this section we look at those with $N=3$ vertices.
\begin{theorem}\label{N=3theorem}
The period $2$ quivers with $3$ nodes are given as follows. For $\sigma=(123)$ and $k=2$:
\begin{equation}\label{N=3result}
\begin{tikzcd}
& 1 \arrow[dl, "n" description]\arrow[dr, "n" description]& \\
2\arrow[rr, "n" description]  & & 3 
\end{tikzcd}
\qquad
\begin{tikzcd}
& 1\arrow[dl, "2" description] & \\
2\arrow[rr, "2" description]  & & 3\arrow[ul, "2" description] 
\end{tikzcd}
\end{equation}
For the other case, with $\sigma=(1)(23)$ and $k=2$, we only find disconnected quivers which we shall ignore.
\begin{proof}
For $\sigma=(123)$ and $k=2$ the equations (\ref{brelations}) are
\[
-b_{2,1}=-b_{3,2}, \qquad -b_{3,1}=b_{1,2}+\epsilon_{1,3,2}, \qquad b_{3,2}+\epsilon_{3,1,2}=-b_{1,3}.
\]
Due to the first equation, without loss of generality, we take $n:=b_{1,2}=b_{2,3}\geq 0$. The second and third then coincide:
\[
n[b_{3,1}]=b_{3,1}+n.
\]
Taking $b_{3,1}\leq 0$ gives the left quiver of (\ref{N=3result}). If $b_{3,1}\geq 0$ then we have $b_{3,1}=\frac{n}{n-1}$ which has the only solution $n=2$, leading to the right quiver of (\ref{N=3result}).
\end{proof}
\end{theorem}
\section{Period $2$ quivers with $N=4$ vertices}
Here we look for period $2$ quivers with $4$ vertices. Due to Lemma \ref{finalformlemma} there are $3$ choices for the pair $\sigma$ and $k$.
\begin{theorem}\label{N=4theorem}
The period $2$ quivers with $4$ nodes are given as follows. For $\sigma=(1,2,3,4)$ and $k=2$:
\begin{equation}\label{first4node}
\begin{tikzcd}
& 1 & \\
2 \arrow[ur, "n" description]& & 4\arrow[ul, "n" description] \\
& 3 \arrow[ul,"n" description]\arrow[ur,"n" description]& 
\end{tikzcd}
\end{equation}
For $\sigma=(1,2)(3,4)$ and $k=3$:
\begin{equation}\label{N=42cyclefive}
\begin{tikzcd}
& 1\arrow[dr, "m" description, bend left]\arrow[dd, "n" description, near end]  & \\
2\arrow[rr, "n" description, near end]\arrow[ur, "l" description, bend left] & & 4\arrow[dl, "l" description, bend left]  \\
& 3\arrow[ul, "m+nl" description, bend left] & 
\end{tikzcd}
\qquad
\begin{tikzcd}
& 1\arrow[dl, "l" description, bend right]\arrow[dr, "m" description, bend left]\arrow[dd, "n" description, near end]  & \\
2\arrow[rr, "n" description, near end] & & 4  \\
& 3\arrow[ul, "m" description, bend left]\arrow[ur, "p" description, bend right] & 
\end{tikzcd}
\end{equation}
For $\sigma=(1)(2,3,4)$ and $k=2$:
\begin{equation}\label{N=42cycleseven}
\begin{tikzcd}
& 1\arrow[dd, "n" description, near end]\arrow[dl, "n" description, bend right]  & \\
2\arrow[rr, "n^2+m" description, near end]\arrow[dr, "m" description, bend right] & & 4\arrow[dl, "m" description, bend left]\arrow[ul, "n" description, bend right]  \\
& 3 & 
\end{tikzcd}
\qquad
\begin{tikzcd}
& 1\arrow[dd, "n" description, near end]\arrow[dl, "n" description, bend right]  & \\
2\arrow[rr, "n^2(m+1)-m" description] & & 4\arrow[ul, "n(m+1)" description, bend right]  \\
& 3\arrow[ur, "m" description, bend right]\arrow[ul, "m" description, bend left] & 
\end{tikzcd}
\end{equation}
\begin{proof}
For $N=4$ the calculations here are not too involved, so we just give the proof for the case $\sigma=(1,2,3,4)$ and $k=2$ as an example. The equations (\ref{brelations}) give 
\begin{empheq}[box=\fbox]{align}
b_{2,1}&=b_{3,2}, & -b_{3,1}&=b_{4,2}+\epsilon_{4,3,2}, & -b_{4,1}&=b_{1,2}+\epsilon_{1,3,2}, \label{boxone} \\
\quad b_{3,2}+\epsilon_{3,1,2}&=-b_{4,3}, & b_{4,2}+\epsilon_{4,1,2}&=-b_{1,3}, & b_{4,3}+\epsilon_{4,1,3}&=b_{1,4}+\epsilon_{1,3,4}.\quad \nonumber
\end{empheq}
Without loss of generality we can take $n:=b_{2,1}=b_{3,2}\geq 0$. In this work we often found it impossible to solve these types of systems without making some assumptions. In this case we first assume that $b_{3,1}\geq 0$. The third equation of \eqref{boxone} then gives $b_{4,1}=n$, since $\epsilon_{1,3,2}=0$ with these assumptions. Similarly the fourth equation gives $b_{3,4}=n$. We now have that $\epsilon_{4,3,2}=0$ so the second equation gives $b_{2,4}=b_{3,1}\geq 0$. At this point we know the directions of all arrows and all equations of \eqref{boxone} are satisfied except for the last, which becomes $nb_{3,1}=0$. If $n=0$ then the quiver we obtain is disconnected, which we shall ignore. Hence $b_{3,1}=0$ so our solution is \eqref{first4node}.

We now return to \eqref{boxone} and instead assume that $b_{1,3}>0$. The third and fourth equations give
\[
b_{4,1}=n(1-b_{1,3}) \leq 0, \qquad b_{3,4}=n(1-b_{1,3}) \leq 0
\]
respectively. We can then subtract the second and fifth equations to get $nb_{1,4}=b_{1,3}$, hence
\[
n^2(b_{1,3}-1)=b_{1,3} \quad \implies \quad n^2=\frac{b_{1,3}}{b_{1,3}-1}
\]
which is an integer only for $b_{1,3}=2$, but then we have $n^2=2$, so there are no solutions in this case.

The solutions for the rest of the permutations and choices for $k$ are found in a similar way.
\end{proof}
\end{theorem}
\section{Period $2$ quivers with $N=5$ vertices for the permutation $\sigma=(1,2,3,4,5)$}
Once we look at the case with $N=5$ the problem becomes much more complicated. In this case Lemma \ref{finalformlemma} gives $4$ choices for the pair $\sigma$ and $k$. In this section we give the solutions and proofs for $\sigma=(1,2,3,4,5)$ and $k=2,3$. 
\begin{theorem}\label{5nodeperiod2theorem}
The period $2$ quivers with $5$ nodes with the permutation $(1,2,3,4,5)$ are given as follows. For $k=2$ we have the $6$ families
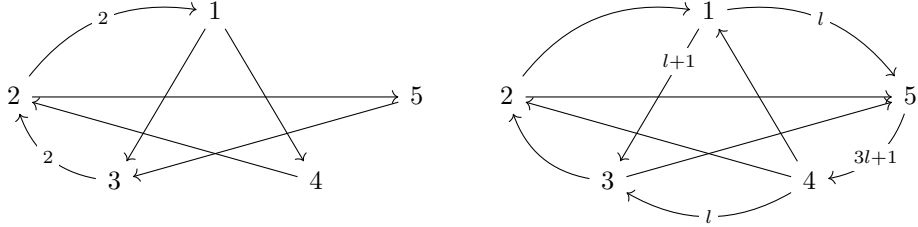
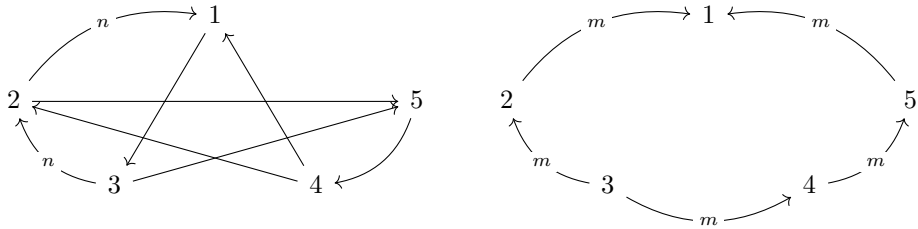
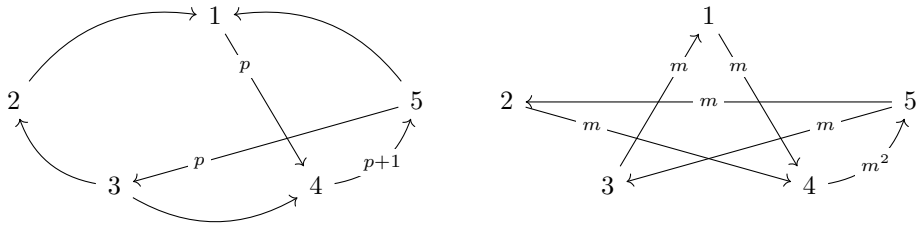
\begin{equation}\label{k=2first}
\begin{tikzcd}
& & 1 \arrow[ddl] \arrow[ddr] & & \\
2 \arrow[rrrr] \arrow[urr, "2" description, bend left]& & & & 5 \arrow[dlll] \\
&3  \arrow[ul, "2" description, bend left]& & 4 \arrow[ulll] &
\end{tikzcd}
\qquad 
\begin{tikzcd}
& & 1 \arrow[ddl,"l+1" description, near start]  \arrow[drr, "l" description, bend left]& & \\
2 \arrow[rrrr, near end] \arrow[urr, bend left]& & & & 5  \arrow[dl, "3l+1" description, bend left]\\
&3\arrow[urrr, near end]  \arrow[ul, bend left]& & 4\arrow[uul, near end] \arrow[ll, "l" description, bend left]\arrow[ulll, near end] &
\end{tikzcd}
\end{equation}
\begin{equation}\label{k=2new}
\begin{tikzcd}
& & 1 \arrow[ddl]  & & \\
2 \arrow[rrrr] \arrow[urr, "n" description, bend left]& & & & 5\arrow[dl, bend left]  \\
&3 \arrow[urrr] \arrow[ul, "n" description, bend left]& & 4\arrow[uul] \arrow[ulll] &
\end{tikzcd}
\qquad
\begin{tikzcd}
& & 1  & & \\
2 \arrow[urr, "m" description, bend left]& & & & 5  \arrow[ull, "m" description, bend right] \\
&3  \arrow[ul, "m" description, bend left]\arrow[rr, "m" description, bend right]& & 4  \arrow[ur, "m" description, bend right]&
\end{tikzcd}
\end{equation}
\begin{equation}\label{k=2second}
\begin{tikzcd}
& & 1  \arrow[ddr, "p" description, near start] & & \\
2 \arrow[urr, bend left]& & & & 5 \arrow[dlll,"p" description, near end]  \arrow[ull, bend right] \\
&3  \arrow[ul, bend left]\arrow[rr, bend right]& & 4  \arrow[ur, "p+1" description, bend right]&
\end{tikzcd}
\qquad
\begin{tikzcd}
& & 1  \arrow[ddr, "m" description, near start] & & \\
2 \arrow[drrr, "m" description, near start]& & & & 5 \arrow[dlll,"m" description, near start] \arrow[llll, "m" description]  \\
&3  \arrow[uur,"m" description, near end]& & 4  \arrow[ur, "m^2" description, bend right]&
\end{tikzcd}
\end{equation}
For $k=3$ we have the $8$ families
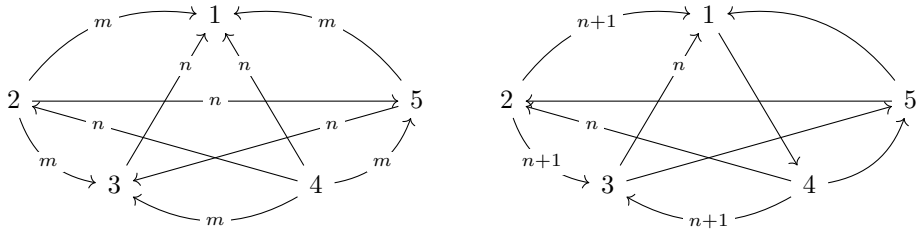
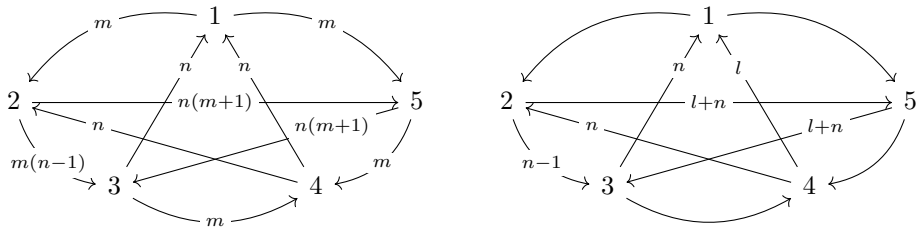
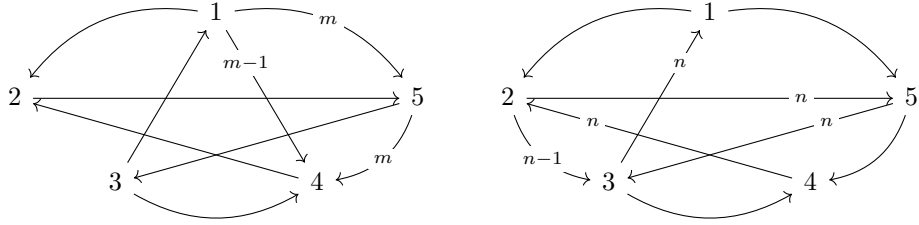
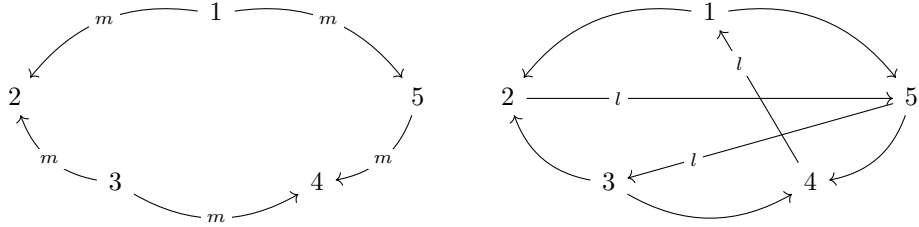
\begin{equation}\label{k=3result}
\begin{tikzcd}
& & 1   & & \\
2\arrow[rrrr, "n" description] \arrow[dr, "m" description, bend right] \arrow[urr, "m" description, bend left]& & & & 5\arrow[ull, "m" description, bend right] \arrow[dlll, "n" description,  near start] \\
& 3 \arrow[uur,"n" description, near end] & & 4 \arrow[ll, "m" description, bend left]\arrow[ur, "m" description, bend right]\arrow[ulll, "n" description, near end] \arrow[uul, "n" description, near end] &
\end{tikzcd}
\qquad
\begin{tikzcd}
& & 1  \arrow[ddr] & & \\
2 \arrow[dr, "n+1" description, bend right] \arrow[urr, "n+1" description, bend left]& & & & 5\arrow[ull, bend right] \arrow[llll] \\
& 3 \arrow[urrr]\arrow[uur,"n" description, near end] & & 4 \arrow[ll, "n+1" description, bend left]\arrow[ur, bend right]\arrow[ulll, "n" description, near end]  &
\end{tikzcd}
\end{equation}
\begin{equation}\label{k=3secondpair}
\begin{tikzcd}
& & 1 \arrow[dll, "m" description, bend right] \arrow[drr, "m" description, bend left] & & \\
2 \arrow[rrrr, "n(m+1)" description]\arrow[dr, "m(n-1)" description, bend right] & & & & 5 \arrow[dlll, "n(m+1)" description, near start]\arrow[dl, "m" description, bend left] \\
& 3 \arrow[rr, "m" description, bend right]\arrow[uur,"n" description, near end] & & 4\arrow[uul, "n" description, near end] \arrow[ulll, "n" description, near end]  &
\end{tikzcd}
\qquad 
\begin{tikzcd}
& & 1 \arrow[dll, bend right] \arrow[drr, bend left] & & \\
2 \arrow[rrrr, "l+n" description]\arrow[dr, "n-1" description, bend right] & & & & 5 \arrow[dlll, "l+n" description, near start]\arrow[dl, bend left] \\
& 3 \arrow[rr, bend right]\arrow[uur,"n" description, near end] & & 4\arrow[uul, "l" description,near end] \arrow[ulll, "n" description, near end]  &
\end{tikzcd}
\end{equation}
\begin{equation}\label{extras}
\begin{tikzcd}
& & 1 \arrow[dll, bend right] \arrow[ddr,"m-1" description, near start]\arrow[drr, "m" description, bend left] & & \\
2 \arrow[rrrr] & & & & 5 \arrow[dlll]\arrow[dl, "m" description, bend left] \\
& 3 \arrow[uur]\arrow[rr, bend right] & & 4 \arrow[ulll]  &
\end{tikzcd}
\qquad
\begin{tikzcd}
& & 1 \arrow[dll, bend right] \arrow[drr, bend left] & & \\
2 \arrow[rrrr, "n" description, near end]\arrow[dr,"n-1" description, bend right] & & & & 5 \arrow[dlll, "n" description, near start]\arrow[dl, bend left] \\
& 3 \arrow[uur,"n" description, near end]\arrow[rr, bend right] & & 4 \arrow[ulll, "n" description, near end]  &
\end{tikzcd}
\end{equation}
\begin{equation}\label{k=3n=0pair}
\begin{tikzcd}
& & 1 \arrow[dll, "m" description, bend right] \arrow[drr, "m" description, bend left] & & \\
2  & & & & 5 \arrow[dl, "m" description, bend left] \\
& 3\arrow[ul, "m" description, bend left] \arrow[rr, "m" description, bend right] & & 4   &
\end{tikzcd}
\qquad 
\begin{tikzcd}
& & 1 \arrow[dll, bend right] \arrow[drr, bend left] & & \\
2 \arrow[rrrr, "l" description, near start] & & & & 5\arrow[dlll, "l" description, near end] \arrow[dl, bend left] \\
& 3\arrow[ul, bend left] \arrow[rr, bend right] & & 4  \arrow[uul, "l" description, near end] &
\end{tikzcd}
\end{equation}
Here mutating the right quiver \eqref{k=3result} at $1$ gives the left quiver \eqref{extras}, up to a relabelling. Similarly mutating the right quiver \eqref{extras} at $1$ gives the right quiver \eqref{k=3n=0pair}, up to a relabelling.
\end{theorem} 
We split the proof in to sections for each $k$, starting with $k=2$.
\subsection{$k=2$}
We assume $n:=b_{3,2}=b_{2,1}> 0$. We mention the case with $n=0$ in Section \ref{letn=0}. Our equations (\ref{brelations}) here are
\begin{empheq}[box=\fbox]{align}
n&=b_{2,1}=b_{5,1}+n[b_{1,3}], & b_{3,1}&=b_{5,2}-n[b_{1,5}], \nonumber \\
\quad b_{4,1}&=b_{3,5}-\epsilon_{5,1,3}+\epsilon_{1,3,4}, & b_{5,1}&=b_{4,5}-\epsilon_{5,1,4}+\epsilon_{1,3,5},\quad \nonumber \\
b_{3,2}&=b_{2,1}, & b_{4,2}&=b_{1,3}-n[b_{4,3}], \label{k=2equations} \\
b_{5,2}&=b_{1,4}-n[b_{5,3}], & b_{4,3}&=-n+n[b_{1,3}], \nonumber \\
b_{5,3}&=b_{2,4}+n[b_{1,4}], & b_{5,4}&=b_{4,3}+\epsilon_{4,1,3}-\epsilon_{5,3,4}. \nonumber
\end{empheq}
These are difficult to solve without further assumptions, so we split into further sections based on the sign of $b_{1,3}$.
\subsubsection{Assuming $b_{1,3}>0$}
Under this assumption the first of (\ref{k=2equations}) is $b_{5,1}=n(1-b_{1,3})$ so we see that $b_{5,1}\leq 0 $. Similarly the eighth equation gives $b_{4,3}=n(b_{1,3}-1)=b_{1,5}\geq 0$ and the tenth gives $b_{5,4}\geq 0 $. The third equation gives $b_{4,1}=b_{3,5}$ which allows us to write the seventh equation as
\[
b_{5,2}=b_{1,4}-n[b_{1,4}]
\]
which, by checking either possible sign for $b_{1,4}$, gives $b_{5,2}\leq0$. Similarly the ninth equation also becomes 
\[
b_{2,4}=b_{1,4}-n[b_{1,4}]\leq 0.
\]
At this point the only sign we have to determine is $b_{1,4}=b_{5,3}$. Firstly we assume $m:=b_{1,4}=b_{5,3}\geq 0$, in this case we arrive at the quiver
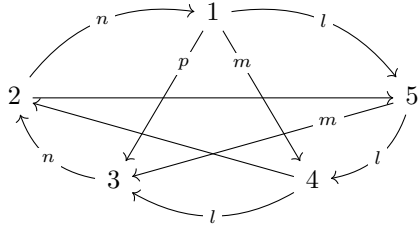
\[
\begin{tikzcd}
& & 1 \arrow[ddl,"p" description, near start] \arrow[ddr, "m" description, near start] \arrow[drr, "l" description, bend left]& & \\
2 \arrow[rrrr, near end] \arrow[urr, "n" description, bend left]& & & & 5 \arrow[dlll,"m" description, near start] \arrow[dl, "l" description, bend left]\\
&3  \arrow[ul, "n" description, bend left]& & 4 \arrow[ll, "l" description, bend left]\arrow[ulll] &
\end{tikzcd}
\]
with $b_{2,5}=b_{4,2}=m(n-1)$ and the relations
\[
l=n(p-1), \qquad p=m(n-1)+nl.
\]
If $p=1$ we get the solution $(l,m,n,p)=(0,1,2,1)$ which is the left (\ref{k=2first}). If $p\neq 1$ we have
\[
n=\frac{l}{p-1}=\frac{p+m}{l+m}
\]
hence $p-1\leq l \leq p$ since $n\in \mathbb{Z}$, but neither $l=p-1$ or $l=p$ gives a solution for $n\neq 0 $. 

We now assume instead that $m:=b_{4,1}=b_{3,5}\geq 0$, where we have the quiver
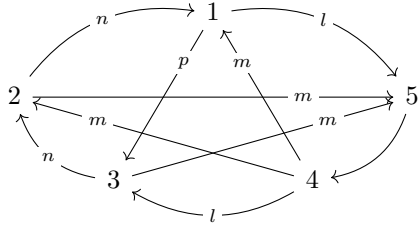
\[
\begin{tikzcd}
& & 1 \arrow[ddl,"p" description, near start]  \arrow[drr, "l" description, bend left]& & \\
2 \arrow[rrrr, "m" description, near end] \arrow[urr, "n" description, bend left]& & & & 5  \arrow[dl, bend left]\\
&3\arrow[urrr,"m" description, near end]  \arrow[ul, "n" description, bend left]& & 4\arrow[uul, "m" description, near end] \arrow[ll, "l" description, bend left]\arrow[ulll, "m" description, near end] &
\end{tikzcd}
\]
with $b_{5,4}=l+lm+mp$ and the relations
\[
n(p-1)=l, \qquad p=m+nl.
\]
Since $m\geq 0$ we have $p\geq nl$. Combining this with the left equation gives
\[
\frac{l}{n}+1\geq nl \qquad \implies \qquad 1\geq l\left(n-\frac{1}{n}\right)\geq 0 
\]
with the solutions $l=n=1$ or $l=0$. These give the right quiver (\ref{k=2first}) and the left quiver \eqref{k=2new}, respectively.
\subsubsection{Assuming $b_{1,3}\leq 0$}
Now, assuming $b_{1,3}\leq 0 $ in (\ref{k=2equations}) we quickly arrive at the quiver
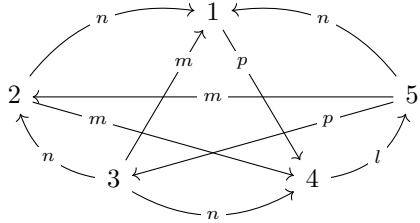
\[
\begin{tikzcd}
& & 1  \arrow[ddr, "p" description, near start] & & \\
2 \arrow[urr, "n" description, bend left]\arrow[drrr, "m" description, near start]& & & & 5 \arrow[dlll,"p" description, near start] \arrow[llll, "m" description] \arrow[ull, "n" description, bend right] \\
&3  \arrow[ul, "n" description, bend left]\arrow[rr, "n" description, bend right]\arrow[uur,"m" description, near end]& & 4  \arrow[ur, "l" description, bend right]&
\end{tikzcd}
\]
with 
\[
n=l-np-mp, \qquad p=m+np.
\]
If $n=1$ then we have the one parameter family of solutions $(l,m,n,p)=(p+1,0,1,p)$, the left (\ref{k=2second}). If $n\neq 1$ then $0\leq p=\frac{m}{1-n}$ hence $p=m=0$ which gives \eqref{k=2new}, right.
\subsubsection{Assuming $n=0$}\label{letn=0}
Assuming $n=0$ in (\ref{k=2equations}) gives the solution (\ref{k=2second}), on the right. 
\subsection{$k=3$}
Without loss of generality we assume $b_{3,1}=b_{4,2}=n> 0$. The case with $n=0$ is left to Section \ref{setn=0}. The equations for each $b_{i,j}$ are
\begin{empheq}[box=\fbox]{align}
\quad b_{2,1}&=b_{5,1}+\epsilon_{2,4,1}=b_{5,1}+n[b_{1,4}], & n&=b_{3,1}=b_{2,5}+\epsilon_{2,1,5}-\epsilon_{3,4,1},\quad \nonumber\\
b_{4,1}&=b_{5,3}+\epsilon_{5,1,3}=b_{5,3}-n[b_{1,5}], & b_{5,1}&=b_{4,5}+\epsilon_{4,1,5}-\epsilon_{5,4,1}, \nonumber \\
b_{3,2}&=-b_{2,1}+\epsilon_{2,4,3}=b_{1,2}-n[b_{3,4}], & b_{4,2}&=b_{3,1}=n, \label{k=3equations}\\
b_{5,2}&=b_{1,4}-n[b_{5,4}], & b_{4,3}&=b_{2,3}-n[b_{1,2}], \nonumber \\
b_{5,3}&=n+\epsilon_{4,1,2}-\epsilon_{5,4,3}, & b_{5,4}&=b_{3,4}+n[b_{1,4}]. \nonumber
\end{empheq}
We first look at the fourth equation, for $b_{4,1}\geq 0$ it becomes
\[
b_{5,1}=b_{4,5}+b_{4,1}[b_{1,5}]-b_{4,1}[b_{5,4}].
\]
Assuming $b_{5,4}\leq 0$ then 
\[
b_{5,1}=b_{4,5}+b_{4,1}[b_{1,5}]\geq 0
\]
so $[b_{1,5}]=0$ and $b_{5,1}=b_{4,5}$. If instead we assume that $b_{5,4}\geq 0$ then 
\[
b_{5,1}=b_{4,5}+b_{4,1}[b_{1,5}]-b_{4,1}b_{5,4}
\]
which cannot hold if $b_{1,5}\leq 0$, so here
\[
b_{5,1}=b_{4,5}+b_{4,1}b_{1,5}-b_{4,1}b_{5,4} \implies b_{5,1}(1+b_{4,1})=b_{4,5}(1+b_{4,1})
\]
hence in this case $b_{5,1}=b_{4,5}$. A similar argument holds if we assume $b_{4,1}\leq 0$, so the fourth equation in (\ref{k=3equations}) is equivalent to $b_{5,1}=b_{4,5}$. The first and last equations together then give $b_{2,1}=b_{4,3}$. The fifth equation is then
\[
b_{3,2}=b_{1,2}-n[b_{1,2}]
\]
which is non-positive for either sign of $b_{1,2}$, since $n>0$. Now, if $b_{4,1}\geq 0$ then 
\[
\epsilon_{3,4,1}=[b_{3,4}]b_{4,1}, \qquad \epsilon_{4,1,2}=b_{4,1}[b_{1,2}]
\]
which are equal. The same hold if $b_{4,1}<0$. Similarly if $b_{2,1}=b_{4,3}\geq 0$ then 
\[
\epsilon_{2,1,5}=b_{2,1}[b_{1,5}], \qquad \epsilon_{5,4,3}=[b_{5,4}]b_{4,3}
\]
which are also equal, and equality also holds for $b_{2,1}<0$. From these we see that the second and ninth equation give $b_{2,5}=b_{5,3}$. At this point our system (\ref{k=3equations}) is reduced to
\begin{empheq}[box=\fbox]{align}
\quad b_{2,1}&=b_{4,3}=b_{5,1}+n[b_{1,4}], & n&=b_{2,5}+\epsilon_{2,1,5}-\epsilon_{3,4,1}, & b_{5,1}&=b_{4,5},\quad \nonumber \\
b_{2,3}&=b_{2,1}+n[b_{1,2}]\geq 0, & b_{5,2}&=b_{1,4}-n[b_{1,5}], & b_{5,3}&=b_{2,5}.\label{k=3reduced}
\end{empheq}
\subsubsection{Assuming $b_{5,1}\geq 0$}
This assumption that $b_{5,1}\geq0$ gives $b_{2,1}=b_{4,3}=b_{2,3}\geq 0$ and $b_{5,2}=b_{1,4}=b_{3,5}$ and a final pair of equations
\[
n=b_{4,1}+[b_{1,4}]b_{2,1}, \qquad b_{2,1}=b_{5,1}+n[b_{1,4}].
\]
With $b_{1,4}\leq0$ the solution appears on the left of (\ref{k=3result}) and with $b_{1,4}>0$ we have
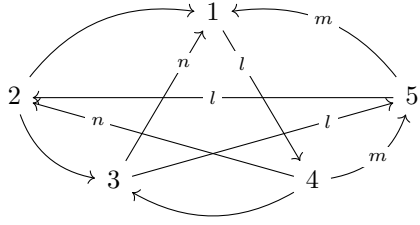
\[
\begin{tikzcd}
& & 1  \arrow[ddr, "l" description, near start] & & \\
2 \arrow[dr, bend right] \arrow[urr, bend left]& & & & 5\arrow[ull, "m" description, bend right] \arrow[llll, "l" description] \\
& 3 \arrow[urrr, "l" description, near end]\arrow[uur,"n" description, near end] & & 4 \arrow[ll, bend left]\arrow[ur, "m" description, bend right]\arrow[ulll, "n" description, near end]  &
\end{tikzcd}
\] 
with $b_{2,1}=b_{2,3}=b_{4,3}=m+nl$ and the relation $n=l(m+nl-1)$. Having $l=1$ gives the solution $(l,m,n)=(1,1,n)$, the right of (\ref{k=3result}), if not we have
\[
n=\frac{l(1-m)}{l^2-1}.
\] 
If $l=0$ then $n=0$, which we have excluded here, hence $l^2-1>0$ so we require $l(1-m)>0$, giving $m=0$. In this case we would then have
\[
n=\frac{l}{l^2-1}
\] 
which has no solutions.
\subsubsection{Assuming $b_{1,5}\geq 0$}
We return to (\ref{k=3reduced}) and assume instead that $b_{1,5}\geq 0$. If $b_{4,1}\geq 0$ then we get two solutions of the form
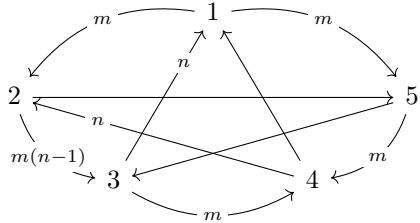
\[
\begin{tikzcd}
& & 1 \arrow[dll, "m" description, bend right] \arrow[drr, "m" description, bend left] & & \\
2 \arrow[rrrr]\arrow[dr, "m(n-1)" description, bend right] & & & & 5 \arrow[dlll]\arrow[dl, "m" description, bend left] \\
& 3 \arrow[rr, "m" description, bend right]\arrow[uur,"n" description, near end] & & 4\arrow[uul] \arrow[ulll, "n" description, near end]  &
\end{tikzcd}
\] 
If $m\neq 1$ then $b_{4,1}=n$ and $b_{2,5}=b_{5,3}=n(1+m)$. For $m=1$ we instead have $b_{2,5}=b_{5,3}=b_{4,1}+n$. This is the pair (\ref{k=3secondpair}). \\\\
If $b_{1,5}\geq 0$, $b_{1,4}\geq 0$ and $b_{1,2}\geq 0$ then the second and sixth equations of (\ref{k=3reduced}) give $b_{2,5}=b_{5,3}=n$. The fifth equation then gives $b_{1,4}=n(b_{1,5}-1)$. Finally the first equation gives 
\[
b_{1,2}=b_{1,5}-nb_{1,4}=b_{1,5}-n^2(b_{1,5}-1).
\] 
Since we are requiring that $b_{1,2}\geq 0$, this has the only solutions $n=1$, $b_{1,5}=0$ or $b_{1,5}=1$, however $b_{1,5}=0$ is not allowed as it will mean $b_{1,4}<0$. The two remaining options give (\ref{extras}). \\\\
Finally for $b_{1,5}\geq0$, $b_{1,4}\geq 0$ and $b_{2,1}\geq 0$ in (\ref{k=3reduced}) we have the equations
\begin{align*}
&b_{2,1}=nb_{1,4}-b_{1,5}, & &n=b_{2,5}+b_{2,1}b_{1,5}+b_{2,1}b_{1,4}, & &b_{5,2}=b_{1,4}-nb_{1,5}, \\
&b_{5,3}=b_{2,5}, & &b_{2,1}=b_{2,3}=b_{4,3}, & &b_{1,5}=b_{5,4}
\end{align*}
with the sign of $b_{2,5}=b_{5,3}$ yet to be determined. The second and third equations together give
\[
0=n(b_{1,5}-1)+b_{1,4}(b_{2,1}-1)+b_{2,1}b_{1,5}.
\]
If the three terms here are all at least $0$ then they are all necessarily $0$, in which case $b_{1,5}=1$ and $b_{2,1}=b_{1,4}=0$, but this doesn't give a solution. If one term is negative then either $b_{1,5}=0$ or $b_{2,1}=0$, both giving no solutions. 
\subsubsection{Assuming $n=0$}\label{setn=0}
Finally returning to (\ref{k=3equations}) and assuming $n=0$ gives the two quivers (\ref{k=3n=0pair}).
\section{Period $2$ quivers with $N=5$ vertices for other permutations}
Here we state the rest of the period two quivers for $N=5$ without proof, since they were obtained using the same methods as for $\sigma=(1,2,3,4,5)$. There are two equations in this case, for $\sigma=(1)(2,3,4,5)$ and $k=2$, and for $\sigma=(1,2)(3,4,5)$ and $k=3$.
\begin{theorem}\label{feniatheorem}
The period $2$ quivers with $N=5$ for permutations other than $(1,2,3,4,5)$ are given as follows. For $(1,2)(3,4,5)$ and $k=3$: 
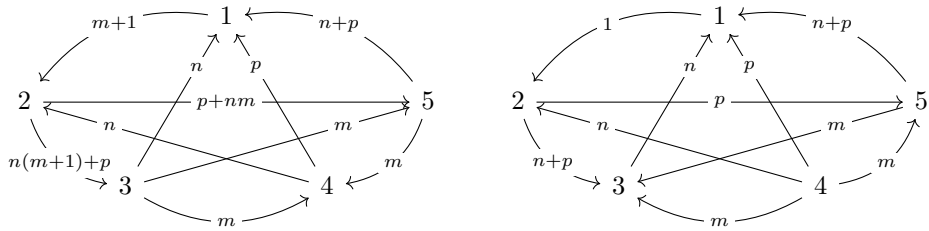
\begin{equation}\label{fenia}
\begin{tikzcd}
& & 1\arrow[dll,"m+1" description, bend right]   & & \\
2\arrow[dr,"n(m+1)+p" description, bend right]\arrow[rrrr,"p+nm" description] & & & & 5\arrow[ull,"n+p" description, bend right]\arrow[dl,"m" description, bend left]   \\
&3\arrow[uur,"n" description, near end]\arrow[urrr,"m" description, near end]\arrow[rr,"m" description, bend right]  & & 4\arrow[uul,"p" description, near end]\arrow[ulll,"n" description, near end]
\end{tikzcd}
\qquad
\begin{tikzcd}
& & 1\arrow[dll,"1" description, bend right]   & & \\
2\arrow[dr,"n+p" description, bend right]\arrow[rrrr,"p" description] & & & & 5\arrow[dlll,"m" description, near start]\arrow[ull,"n+p" description, bend right]   \\
&3\arrow[uur,"n" description, near end]  & & 4\arrow[ulll,"n" description, near end]\arrow[uul,"p" description, near end]\arrow[ur,"m" description, bend right]\arrow[ll,"m" description, bend left]
\end{tikzcd}
\end{equation}
For $(1)(2,3,4,5)$ and $k=2$:
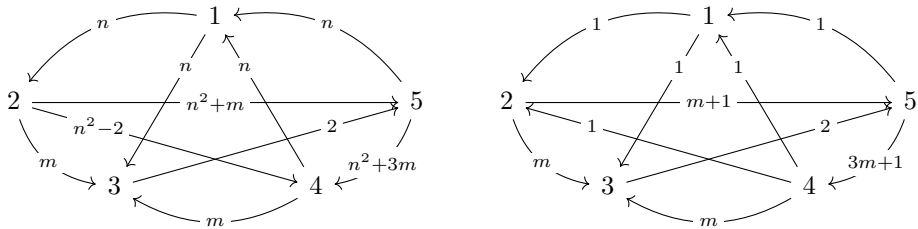
\begin{equation}\label{fenia2}
\begin{tikzcd}
& & 1\arrow[dll,"n" description, bend right]\arrow[ddl,"n" description, near start]  & & \\
2\arrow[dr,"m" description, bend right]\arrow[drrr,"n^2-2" description, near start]\arrow[rrrr,"n^2+m" description] & & & & 5\arrow[ull,"n" description, bend right]\arrow[dl,"n^2+3m" description, bend left]   \\
&3\arrow[urrr,"2" description, near end]  & & 4\arrow[uul,"n" description, near end]\arrow[ll,"m" description, bend left]
\end{tikzcd}
\qquad
\begin{tikzcd}
& & 1\arrow[dll,"1" description, bend right]\arrow[ddl,"1" description, near start]  & & \\
2\arrow[dr,"m" description, bend right]\arrow[rrrr,"m+1" description] & & & & 5\arrow[ull,"1" description, bend right]\arrow[dl,"3m+1" description, bend left]   \\
&3\arrow[urrr,"2" description, near end]  & & 4\arrow[ulll,"1" description, near end]\arrow[uul,"1" description, near end]\arrow[ll,"m" description, bend left]
\end{tikzcd}
\end{equation}
where the left quiver requires $n>1$. We also have
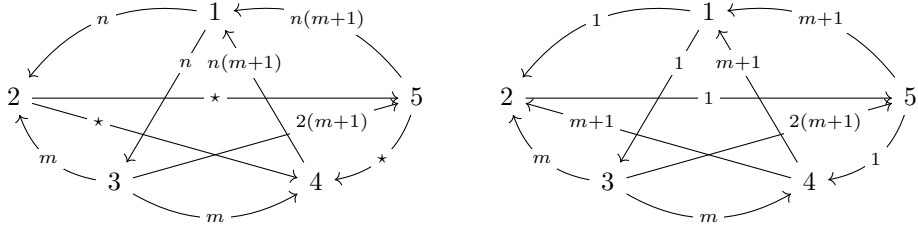
\begin{equation}\label{fenia3}
\begin{tikzcd}
& & 1\arrow[dll,"n" description, bend right]\arrow[ddl,"n" description, near start]   & & \\
2\arrow[drrr, "\star" description, near start]\arrow[rrrr, "\star" description] & & & & 5\arrow[ull,"n(m+1)" description, bend right]\arrow[dl, "\star" description, bend left]   \\
&3\arrow[urrr,"2(m+1)" description, near end]\arrow[ul,"m" description, bend left]\arrow[rr,"m" description, bend right]  & & 4\arrow[uul,"n(m+1)" description, near end]
\end{tikzcd}
\qquad
\begin{tikzcd}
& & 1\arrow[dll,"1" description, bend right]\arrow[ddl,"1" description, near start]   & & \\
2\arrow[rrrr, "1" description] & & & & 5\arrow[ull,"m+1" description, bend right]\arrow[dl, "1" description, bend left]   \\
&3\arrow[urrr,"2(m+1)" description, near end]\arrow[ul,"m" description, bend left]\arrow[rr,"m" description, bend right]  & & 4\arrow[ulll, "m+1" description, near end]\arrow[uul,"m+1" description, near end]
\end{tikzcd}
\end{equation}
with $b_{2,5}=b_{5,4}=n^2(m+1)-m$ and $b_{2,4}=(n^2-2)(m+1)$ in the left quiver, where we need to take $n>1$.
\end{theorem}
\section{Period $2$ quivers with $N=6$, for $k=5$ and the permutation $(1,2,3,4,5,6)$}
For $6$ node quivers, finding all solutions to the period $2$ equation becomes very involved, so in this section we find the $6$ node period $2$ quivers only for the specific choice $k=5$ and $\sigma=(1,2,\ldots,6)$. We note that, in order to fit in with our assumptions in Lemma \ref{finalformlemma} we ought to relabel these solutions so that $k\rightarrow 2$. 
\begin{theorem}\label{N=6theorem}
The period $2$ quivers with $N=6$ and $k=5$ with the permutation $(1,2,3,4,5,6)$ are precisely
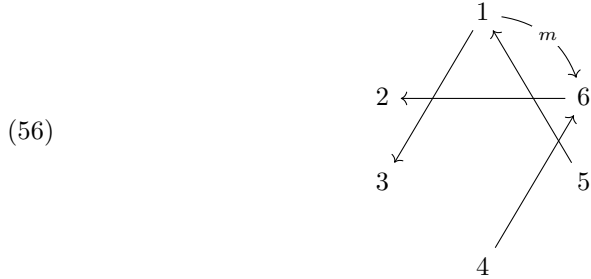
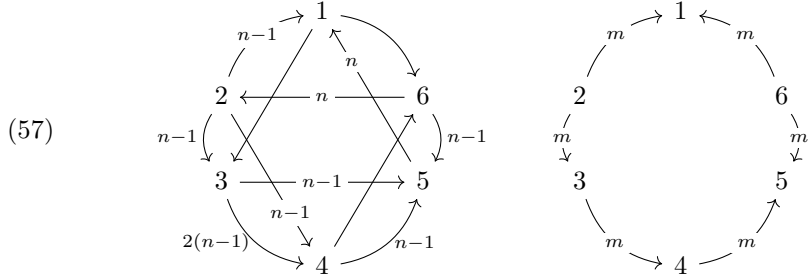
\begin{equation}\label{N=6result}
\begin{tikzcd}
& 1\arrow[ddl] \arrow[dr, "m" description, bend left] & \\
2 & & 6  \arrow[ll]  \\
3 & & 5 \arrow[uul] \\
& 4 \arrow[uur]  &
\end{tikzcd}
\end{equation}
\begin{equation}\label{N=6secondset}
\begin{tikzcd}
& 1\arrow[ddl]\arrow[dr, bend left] & \\
2\arrow[d, "n-1" left, bend right] \arrow[ur, "n-1" description, bend left]\arrow[ddr, "n-1" description, near end]& & 6\arrow[d,"n-1" right, bend left]  \arrow[ll, "n" description]  \\
3 \arrow[dr, "2(n-1)" left, bend right] \arrow[rr, "n-1" description] & & 5 \arrow[uul, "n" description, near end] \\
& 4 \arrow[uur] \arrow[ur, "n-1" right, bend right]&
\end{tikzcd}
\qquad 
\begin{tikzcd}
& 1 & \\
2 \arrow[d, "m" description, bend right]\arrow[ur, "m" description, bend left]& & 6 \arrow[ul, "m" description, bend right] \arrow[d, "m" description, bend left] \\
3 \arrow[dr, "m" description, bend right] & & 5 \\
& 4 \arrow[ur, "m" description, bend right] &
\end{tikzcd}
\end{equation}
\end{theorem}
We take $n:=b_{5,1}=b_{6,2}>0$ and leave $n=0$ until the end of this section. In this case (\ref{brelations}) gives $15$ equations, which we shall not write down before noting some simplifications. Firstly we have
\[
b_{1,6}=b_{5,6}+n[b_{1,6}], \qquad b_{1,2}=b_{1,6}-n[b_{1,6}]
\]
which give $b_{6,5}\geq 0$ and $b_{1,2}\leq 0$ respectively. Secondly
\[
b_{3,2}=b_{1,2}-n[b_{3,6}], \qquad b_{6,5}=b_{4,5}-n[b_{1,4}]
\]
so $b_{3,2}\leq 0$ and $b_{4,5}\geq 0$. At this point we have the quiver
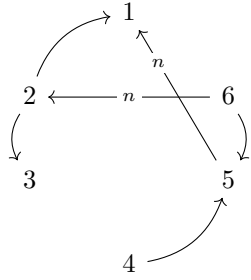
\[
\begin{tikzcd}
& 1 & \\
2 \arrow[d, bend right]\arrow[ur, bend left]& & 6 \arrow[ll, "n" description] \arrow[d, bend left] \\
3 & & 5 \arrow[uul, "n" description, near end]\\
& 4 \arrow[ur, bend right] &
\end{tikzcd}
\]
with the orientations of the undrawn arrows still to be determined, and the relations 
\begin{empheq}[box=\fbox]{align}
b_{2,1}&=n[b_{1,6}]-b_{1,6}, & b_{1,3}&=n-b_{2,1}[b_{1,6}]-\epsilon_{1,6,3}, \nonumber\\
b_{1,4}&=b_{6,3}+\epsilon_{6,1,3}-\epsilon_{1,6,4}, & b_{4,6}&=n+\epsilon_{6,1,4}-[b_{1,6}]b_{6,5}, \nonumber \\
b_{6,5}&=b_{6,1}+n[b_{1,6}], & b_{2,3}&=b_{2,1}+n[b_{3,6}], \nonumber \\
\quad b_{4,2}&=b_{1,3}-n[b_{4,6}], & b_{5,2}&=b_{1,4}, \label{N=6relations} \\
b_{3,4}&=b_{2,3}+b_{2,1}[b_{1,3}]-\epsilon_{3,6,4}, & b_{5,3}&=b_{4,2}-b_{2,1}[b_{1,4}]+[b_{3,6}]b_{6,5},\quad \nonumber \\
b_{6,3}&=b_{2,5}, & b_{4,5}&=b_{3,4}+\epsilon_{3,1,4}-[b_{4,6}]b_{6,5}, \nonumber \\
b_{6,4}&=b_{3,5}-n[b_{1,3}], & b_{6,5}&=b_{4,5}-n[b_{1,4}]. \nonumber
\end{empheq}
\subsection{Assuming $b_{1,6}\leq 0$}
With the assumption $b_{1,6} \leq 0$ we see the second and fourth of (\ref{N=6relations}) become
\[
b_{1,3}=n+b_{6,1}[b_{3,6}], \qquad b_{4,6}=n+b_{6,1}[b_{1,4}]
\]
so $b_{1,3}\geq 0$ and $b_{4,6} \geq 0$. In fact, since $b_{1,4}=b_{5,2}=b_{3,6}$ we see that $b_{1,3}=b_{4,6}$. The third equation then gives
\[
2b_{1,4}=b_{6,1}b_{1,3}+b_{4,6}b_{6,1}\geq 0
\]
and the ninth equation
\[
b_{3,4}=b_{2,3}+b_{2,1}[b_{1,3}]+b_{4,6}[b_{6,3}]\geq 0.
\]
Since $b_{1,3}=b_{4,6}$, the seventh and thirteenth of (\ref{N=6relations}) give
\[
b_{2,4}=b_{3,5}=b_{1,3}(n-1)\geq 0.
\]
Now we have the direction of every arrow, our quiver looks like 
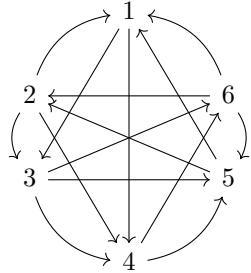
\[
\begin{tikzcd}
& 1 \arrow[ddd] \arrow[ddl] & \\
2 \arrow[d, bend right]\arrow[ur, bend left]\arrow[ddr]& & 6  \arrow[ul, bend right]\arrow[ll] \arrow[d, bend left] \\
3\arrow[rr]\arrow[urr]\arrow[dr, bend right] & & 5 \arrow[ull]\arrow[uul]\\
& 4 \arrow[ur, bend right] \arrow[uur] &
\end{tikzcd}
\]
We define $p:=b_{1,3}=b_{4,6}$ and $m:=b_{2,1}=b_{6,1}=b_{6,5}$ so our set (\ref{N=6relations}) becomes 
\begin{align*}
&mp=b_{1,4}=b_{5,2}=b_{3,6}, & &b_{2,3}=b_{4,5}=m(1+np), \\
&b_{3,4}=m(1+p+np), & &b_{2,4}=b_{3,5}=p(n-1),
\end{align*}
with the relation
\[
p=n+m^2p,
\]
however this is only solved by $m=0$ and $p=n$, which gives a disconnected quiver.
\subsection{Assuming $b_{1,6}>0$}
Assuming $b_{1,6}>0$ gives the partially completed quiver 
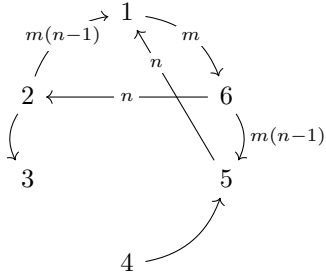
\[
\begin{tikzcd}
& 1 \arrow[dr, "m" description, bend left] & \\
2 \arrow[d, bend right]\arrow[ur, "m(n-1)" description, bend left]& & 6  \arrow[ll, "n" description] \arrow[d, "m(n-1)" right, bend left] \\
3 & & 5 \arrow[uul ,"n" description, near end]\\
& 4 \arrow[ur, bend right]  &
\end{tikzcd}
\]
where we have taken $m:=b_{1,6}$. From (\ref{N=6relations}) we have $b_{5,2}=b_{1,4}=b_{3,6}$ so the third equation there gives
\[
2b_{1,4}=\epsilon_{6,1,3}-\epsilon_{1,6,4}=-m([b_{3,1}]+[b_{6,4}]\leq 0.
\] 
From the twelfth equation of (\ref{N=6relations}) we have 
\[
b_{3,4}=b_{4,5}+[b_{4,6}]b_{6,5}+b_{4,1}[b_{1,3}]\geq 0.
\]
Our updated, but still incomplete, quiver is
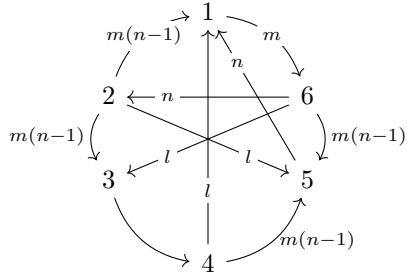
\[
\begin{tikzcd}
& 1 \arrow[dr, "m" description, bend left] & \\
2 \arrow[drr,"l" description, near end]\arrow[d, "m(n-1)" left, bend right]\arrow[ur, "m(n-1)" description, bend left]& & 6 \arrow[dll,"l" description, near end]  \arrow[ll, "n" description, near end] \arrow[d, "m(n-1)" right, bend left] \\
3\arrow[dr, bend right] & & 5 \arrow[uul ,"n" description, near end]\\
& 4 \arrow[uuu ,"l" description, near start] \arrow[ur, "m(n-1)" right, bend right]  &
\end{tikzcd}
\]
where we have defined $l:=b_{4,1}=b_{6,3}=b_{2,5}$. Our remaining relations to satisfy are
\begin{empheq}[box=\fbox]{align}
b_{1,3}&=n-m((n-1)m+l), & 2l&=m([b_{3,1}]+[b_{6,4}]),\quad \nonumber\\
b_{4,6}&=n-m(l+m(n-1)), & b_{4,2}&=b_{1,3}-n[b_{4,6}], \nonumber\\
\quad b_{3,4}&=m(n-1)([b_{1,3}]+1)+l[b_{4,6}], & b_{5,3}&=b_{4,2}, \label{N=6secondcase}\\
b_{3,4}&=m(n-1)(1+[b_{4,6}])+l[b_{1,3}], & b_{6,4}&=b_{3,5}-n[b_{1,3}].\nonumber
\end{empheq}
The first and third give $b_{1,3}=b_{4,6}$ so the second equation gives $l=m[b_{3,1}]$. This also implies, using the fourth equation,
\[
b_{4,2}=b_{1,3}-n[b_{1,3}]\leq 0.
\]
If $b_{3,1}\leq 0 $ then $l=0$ and the first equation is equivalent to
\[
b_{1,3}=n-m^2(n-1)\geq 0
\]
which is satisfied by $n=1$ or $m=0$ or $m=1$. Taking $n=1$ gives (\ref{N=6result}). Taking $m=0$ gives another disconnected quiver and taking $m=1$ gives the left of (\ref{N=6secondset}).\\\\
If we return to (\ref{N=6secondcase}) and take $p:=b_{3,1}\geq 0$ then we again have $l=m[b_{3,1}]=mp$ so the first equation of (\ref{N=6secondcase}) implies
\[
(n+p)(m^2-1)=m^2
\]
which doesn't have $m=1$ as a solution, so
\[
n+p=\frac{m^2}{m^2-1}=1+\frac{1}{m^2-1}
\]
which is an integer only for $m=0$, but in this case $n=p=0$, which we have excluded.
\subsection{Assuming $n=0$}
Finally in (\ref{N=6relations}) taking $n=0$ gives the right of (\ref{N=6secondset}).
\section{$T$- and $Y$-systems for our period $2$ quivers}
In this section we use Nakanishi's formulae to write down some of the more interesting $T$- and $Y$-systems for our period $2$ quivers, as given in Lemmas \ref{firstTsystem} and \ref{secondTsystemlemma} and Section \ref{Ysystemssubsection}. We then consider periodic quantities and how they can be used to simplify these $T$- and $Y$-systems, in some cases giving a single recurrence (the particular systems appearing here were chosen because they exhibit periodic quantities). We compile the results of this section in the following proposition.
\begin{proposition}
The $T$- and $Y$-systems we study in this section have the following properties.
\begin{enumerate}[(i)]
\item The quiver (\ref{first4node}): Both the $T$- and $Y$-system have a period $2$ quantity which can be used to reduce each of the systems to a recurrence.
\item The left quiver \eqref{k=2second}: The $T$-system has a period $1$ quantity so we can write the system as a recurrence, a special case of which is the Somos-$4$ recurrence.
\item The right quiver \eqref{k=3result}: The $T$-system has a period $1$ quantity but it doesn't seem possible to write this system as a recurrence.
\item The right quiver \eqref{k=3n=0pair}: The $T$-system has a period $1$ quantity and the system may be written as the same recurrence (that generalises Somos-$4$) as in part (ii).
\item The quiver \eqref{N=6result}: The $T$-system has a period $2$ quantity that leads to a recurrence.
\item The left quiver \eqref{N=6secondset}: The $T$-system has a period $1$ quantity that leads to a recurrence that has the Somos-$5$ recurrence as a special case.
\end{enumerate}
\end{proposition}
We will discuss each of these quivers in a separate subsection.
\subsection{The quiver (\ref{first4node})}
The $T$-system for this quiver is
\begin{align}
z(q)y(q+1)&=z(q+1)^ny(q)^n+1, \nonumber \\
y(q)z(q+3)&=z(q+2)^ny(q+1)^n+1 \label{firstT}
\end{align}
We can eliminate the terms on the right hand sides to find that 
\[
C(q):=\frac{y(q)}{z(q+1)}
\]
is period $2$. We can use this to replace the $y$ terms in the $T$-system, so that it becomes
\[
C(q+1)z(q+2)z(q)=C(q)^n z(q+1)^{2n}+1.
\]
The corresponding $Y$-system for this quiver is  
\begin{align}
A(q)B(q+1)&=(1+A(q+1))^n(1+B(q))^n, \nonumber \\
B(q)A(q+3)&=(1+A(q+2))^n(1+B(q+1))^n. \label{firstYaugust}
\end{align}
By a similar argument this has 
\[
D(q):=\frac{A(q+1)}{B(q)}
\]
which is period $2$. By replacing the $B$ terms this gives the equation
\begin{equation}\label{firstYreduction}
D(q+1)A(q+2)A(q)=(1+A(q+1))^n(1+D(q)A(q+1))^n.
\end{equation}
\begin{remark}
In this case the $T_Z$-system (see Remark \ref{TZremark}) is given by multiplying the right hand sides of the $T$-system \eqref{firstT} by new variables which we write as $Z_z(q)$ and $Z_y(q)$:
\begin{align}
z(q)y(q+1)&=Z_z(q)(z(q+1)^ny(q)^n+1), \nonumber \\
y(q)z(q+3)&=Z_y(q)(z(q+2)^ny(q+1)^n+1) \nonumber
\end{align}
and the theorem of \cite{honeinoue} says that taking 
\[
\bar{A}(q):=z(q+1)^ny(q)^n, \qquad \bar{B}(q):=z(q+2)^ny(q+1)^n
\]
gives a solution of the $Y$-system \eqref{firstYaugust} if and only if the $Z$ variables satisfy 
\[
Z_z(q+1)^nZ_y(q)^n=Z_z(q+2)^nZ_y(q+1)^n=1.
\]
Here we note that, since $\bar{B}(q)=\bar{A}(q+1)$ this can't give all solutions of the $Y$-system, regardless of any conditions on the $Z$ variables. 

We also note that $C(q)$ only remains period $2$ in the $T_Z$-system if we have $Z_z(q+1)=Z_y(q)$. Similar results hold for the other periodic quantities considered in this section: they are only periodic in the $T_Z$-systems subject to further relations on the $Z$ variables.
\end{remark}
\subsection{The left quiver \eqref{k=2second}}\label{plant}
The $T$-system here is
\begin{align}
z(q)y(q+1)&=z(q+1)y(q)+z(q+2)^p \nonumber \\
y(q)z(q+4)&=z(q+2)^p+z(q+3)y(q+1) \label{firstsystem}
\end{align}
By eliminating the $z(q+2)^p$ terms and factoring, we find that
\[
C:=\frac{z(q)+z(q+3)}{y(q)}
\]
is period $1$. We can then use $C$ to replace each of the $y$ terms in (\ref{firstsystem}), so the $T$-system becomes
\begin{equation}\label{hone}
z(q)z(q+4)=z(q+1)z(q+3)+Cz(q+2)^p.
\end{equation}
We note that with $p=2$ this equation is known as the Somos-4 recurrence and appears in \cite{hone2005elliptic}, where it is solved using the Weierstrass sigma function. It was then noted in \cite{fordyhone} that this recurrence can be obtained from a period $1$ quiver. The associated $Y$-system was shown to be related to the $q$-Painlev\'e I equation in \cite{okubo2013discrete}.
\subsection{The right quiver \eqref{k=3result}}
Here we have the $T$-system
\begin{align}
z(q)y(q+2)&=z(q+1)y(q)^ny(q+1)^{n+1}+z(q+2), \nonumber \\
y(q)z(q+3)&=z(q+1)+z(q+2)y(q+1)^{n+1}y(q+2)^n \label{canreduce}
\end{align}
By shifting the first equation $q\mapsto q+1$ and eliminating the term ${z(q+2)y(q+1)^ny(q+2)^n}$ that is present in both equations, we can then factor the result to show that
\[
\frac{y(q+1)y(q+3)+y(q+2)}{z(q+3)}=\frac{y(q)y(q+2)+y(q+1)}{z(q+1)}
\] 
Multiplying this expression by $1/z(q+2)$ proves that
\[
C:=\frac{y(q)y(q+2)+y(q+1)}{z(q+1)z(q+2)}
\]
is constant. We then multiply the first $T$-system equation by $z(q+1)$ and use $C$ to replace the terms $z(q)z(q+1)$ and $z(q+1)z(q+2)$ that appear to prove that this $T$-system is equivalent to
\begin{align}
y(q+3)y(q)&=Cz(q+2)^2y(q+1)^ny(q+2)^n+1, \nonumber \\
Cz(q+2)z(q+1)&=y(q)y(q+2)+y(q+1). \nonumber
\end{align}
Unfortunately we are not able to remove the $z$ terms completely in the first equation.
\subsection{The right quiver \eqref{k=3n=0pair}}
Here the $T$-system is
\begin{align}
z(q)y(q+2)&=z(q+1)y(q+1)+z(q+2)^l, \nonumber \\
y(q)z(q+3)&=z(q+1)^l+z(q+2)y(q+1). \label{ooh}
\end{align}
By shifting the second equation and eliminating the $z(q+2)^l$ terms we see that
\[
C:=\frac{z(q)+z(q+3)}{y(q+1)}
\]
is constant, which we can then use to replace each of the $y$ terms in (\ref{ooh}). Hence this $T$-system becomes
\[
z(q)z(q+4)=z(q+1)z(q+3)+Cz(q+2)^l
\]
with is another appearance of \eqref{hone}.
\subsection{The quiver \eqref{N=6result}}
We have the $T$-system
\begin{align}
z(q)y(q+4)&=z(q+1)^my(q+2)+y(q), \nonumber \\
y(q)z(q+2)&=z(q+1)^my(q+2)+y(q+4) \label{can}
\end{align}
Subtracting the two equations gives 
\[
\frac{z(q+2)+1}{y(q+4)}=\frac{z(q)+1}{y(q)}
\]
We multiply both sides by $1/y(q+2)$ to show that 
\[
C(q):=\frac{z(q)+1}{y(q+2)y(q)}
\]
is period $2$. We can use this to replace the $z$ terms in \eqref{can} to obtain the recurrence
\begin{align}
C(q)y(q+4)y(q+2)y(q)=(C(q+1)y(q+3)y(q+1)-1)^m+y(q+4)+y(q) \nonumber
\end{align}
\subsection{The left quiver \eqref{N=6secondset}}
The $T$-system here is
\begin{align}
z(q)y(q+4)&=z(q+1)y(q+2)+y(q)^ny(q+3)^{n-1}, \nonumber
\\
y(q)z(q+2)&=z(q+1)y(q+2)+y(q+1)^{n-1}y(q+4)^n \label{waits}
\end{align}
With a shift we have terms $y(q+1)^{n-1}y(q+4)^{n-1}$ appearing on both right hand sides. Eliminating these proves that
\[
C:=\frac{y(q)y(q+1)+y(q+3)y(q+4)}{z(q+1)}
\]
is constant. We can then replace the $z$ terms in the $T$-system to give the recurrence
\begin{equation}\label{okubo}
y(q+5)y(q)=y(q+3)y(q+2)+Cy(q+1)^{n-1}y(q+4)^{n-1}.
\end{equation}
For $n=2$ this is known as the Somos-5 recurrence and was obtained through mutation of a period $1$ quiver in \cite{fordymarsh}. In \cite{hone2007sigma} Somos-5 was solved using Weierstrass sigma functions. In \cite{okubo2013discrete} it was shown that the associated $Y$-system for this $T$-system (with $n=2$) gives rise to the $q$-Painlev\'e II equation.
\section{Concluding Remarks}
We began by justifying why we take the equation $\sigma\mu_k\mu_1(Q)=Q$ as the definition for period $2$ quivers, where we only allow the $2$ permutations $\sigma$ and the vertices $k$ given in Lemma \ref{finalformlemma}. This leads to the complicated system of equations \eqref{brelations}, which we were able to solve for $N=3,4,5$ and, in some cases, $N=6$. We then considered some of the $T$- and $Y$-systems obtained from these quivers which exhibit periodic quantities that can be used to simplify these systems.

In \cite{fordyhone} the authors define a ``cluster map" for period $1$ quivers. This is the map
\[
\varphi:(x_n,\ldots, x_{n+N-1}) \mapsto (x_{n+1},\ldots,x_{n+N})
\]
where $x_{n+N}$ is defined by \eqref{period1reccurence}. There it is proved that, if the matrix $B$ is degenerate then there exists a projection $\pi$ to a set of reduced variables, given by the image of $B$, and a map $\hat{\varphi}$ such that $\pi \circ \varphi= \hat{\varphi} \circ \pi$. The map $\hat{\varphi}$ gives a recurrence called the $U_Z$-system. In \cite{honeinoue} it is then shown that, for certain period $1$ quivers, the $U_Z$-systems are linked to $q$-Painlev\'{e} systems given in \cite{kruskal2000asymmetric}. A similar connection is found in \cite{okubo} where, among other equations, the discrete KdV and the discrete mKdV are constructed as $Y$-systems by mutating quivers with an infinite number of vertices, reductions of these then give rise to $q$-Painlev\'{e} I, II, III and VI. It would be interesting to see if any of the $Y$-systems from period $2$ quivers are related to known systems, possibly the forms of discrete Painlev\'{e} that appear as systems of two equations in \cite{kruskal2000asymmetric}. 

We reiterate that we did not look at the $T$- and $Y$-systems for all of the period $2$ quivers found here, just the ones with obvious periodic quantities. We also remark that these periodic quantities only exist in the $T_Z$ systems subject to certain relations between the $Z$ variables, in many (but not all) cases only if each of the $Z$ variables is equal to $1$. Due to this we are so far unable to say much about the period $2$ $T_Z$- and $Y$-systems beyond what was noted in Remark \ref{TZremark}. Investigating the rest of the period $2$ systems (those without periodic quantities) and the relations between solutions of the $T_Z$- and $Y$-systems is a problem we intend to tackle in the future.
\section*{Acknowledgements}
The author thanks Rei Inoue for helpful discussions and advice. This research was carried out while the author was a recipient of a Japan Society for the Promotion of Science (JSPS) postdoctoral fellowship and was supported by JSPS KAKENHI Grant Number 21F20788.
\bibliographystyle{plain}
\bibliography{Period_Two_Quivers_Update}

\begin{thebibliography}{10}

\bibitem{frises}
Ibrahim {Assem}, Christophe {Reutenauer}, and David {Smith}.
\newblock {Friezes.}
\newblock {\em {Adv. Math.}}, 225(6):3134--3165, 2010.

\bibitem{clusteri}
Sergey Fomin and Andrei Zelevinsky.
\newblock Cluster algebras {I}: Foundations.
\newblock {\em Journal of the American Mathematical Society}, 15(2):497--529,
  2002.

\bibitem{clusterii}
Sergey Fomin and Andrei Zelevinsky.
\newblock Cluster algebras {II}: Finite type classification.
\newblock {\em Inventiones Mathematicae}, 154(1):63--121, 2003.

\bibitem{Ysystemsassociahedra}
Sergey Fomin and Andrei Zelevinsky.
\newblock Y-systems and generalized associahedra.
\newblock {\em Annals of Mathematics}, 158(3):977--1018, 2003.

\bibitem{clustersiv}
Sergey Fomin and Andrei Zelevinsky.
\newblock Cluster algebras {IV}: coefficients.
\newblock {\em Compositio Mathematica}, 143(1):112--164, 2007.

\bibitem{fordyhone}
Allan~P Fordy and Andrew Hone.
\newblock Discrete integrable systems and {Poisson} algebras from cluster maps.
\newblock {\em Communications in Mathematical Physics}, 325(2):527--584, 2014.

\bibitem{fordymarsh}
Allan~P Fordy and Bethany~R Marsh.
\newblock Cluster mutation-periodic quivers and associated {Laurent} sequences.
\newblock {\em Journal of Algebraic Combinatorics}, 34(1):19--66, 2011.

\bibitem{galashin}
Pavel Galashin and Pavlo Pylyavskyy.
\newblock The classification of {Zamolodchikov} periodic quivers.
\newblock {\em American Journal of Mathematics}, 141(2):447--484, 2019.

\bibitem{hone2007sigma}
Andrew Hone.
\newblock Sigma function solution of the initial value problem for {Somos} 5
  sequences.
\newblock {\em Transactions of the American Mathematical Society},
  359(10):5019--5034, 2007.

\bibitem{hone2005elliptic}
Andrew~NW Hone.
\newblock Elliptic curves and quadratic recurrence sequences.
\newblock {\em Bulletin of the London Mathematical Society}, 37(2):161--171,
  2005.

\bibitem{honeinoue}
Andrew~NW Hone and Rei Inoue.
\newblock Discrete {Painlev{\'e}} equations from {Y}-systems.
\newblock {\em Journal of Physics A: Mathematical and Theoretical},
  47(47):474007, 2014.

\bibitem{inoue2010periodicities}
Rei Inoue, Osamu Iyama, Atsuo Kuniba, Tomoki Nakanishi, and Junji Suzuki.
\newblock Periodicities of {T}-systems and {Y}-systems.
\newblock {\em Nagoya Mathematical Journal}, 197:59--174, 2010.

\bibitem{inouenakanishi}
Rei Inoue and Tomoki Nakanishi.
\newblock Difference equations and cluster algebras {I}: {Poisson} bracket for
  integrable difference equations.
\newblock {\em Trans. Amer. Math. Soc}, 362:859--895, 2010.

\bibitem{kedem}
Rinat Kedem.
\newblock Q-systems as cluster algebras.
\newblock {\em Journal of Physics A: Mathematical and Theoretical},
  41(19):194011, 2008.

\bibitem{kellerperiodicity}
Bernhard Keller.
\newblock The periodicity conjecture for pairs of {Dynkin} diagrams.
\newblock {\em Annals of Mathematics}, pages 111--170, 2013.

\bibitem{kellerscherotzke}
Bernhard Keller and Sarah Scherotzke.
\newblock Linear recurrence relations for cluster variables of affine quivers.
\newblock {\em Advances in Mathematics}, 228(3):1842--1862, 2011.

\bibitem{kruskal2000asymmetric}
Martin Kruskal, KM~Tamizhmani, Basile Grammaticos, and Alfred Ramani.
\newblock Asymmetric discrete {Painlev{\'e}} equations.
\newblock {\em Regular and Chaotic Dynamics}, 5(3):273--280, 2000.

\bibitem{kuniba2011review}
Atsuo Kuniba, Tomoki Nakanishi, and Junji Suzuki.
\newblock {T}-systems and {Y}-systems in integrable systems.
\newblock {\em Journal of Physics A: Mathematical and Theoretical},
  44(10):103001, 2011.

\bibitem{mizuno}
Yuma Mizuno.
\newblock Difference equations arising from cluster algebras.
\newblock {\em Journal of Algebraic Combinatorics}, pages 1--57, 2020.

\bibitem{nakanishi}
Tomoki Nakanishi.
\newblock Periodicities in cluster algebras and dilogarithm identities.
\newblock {\em Representations of algebras and related topics}, 5:407, 2011.

\bibitem{okubo2013discrete}
Naoto Okubo.
\newblock Discrete integrable systems and cluster algebras (the breadth and
  depth of nonlinear discrete integrable systems).
\newblock {\em RIMS Kokyuroku Bessatsu}, 41:25--41, 2013.

\bibitem{okubo}
Naoto Okubo.
\newblock Bilinear equations and q-discrete {Painlev{\'e}} equations satisfied
  by variables and coefficients in cluster algebras.
\newblock {\em Journal of Physics A: Mathematical and Theoretical},
  48(35):355201, 2015.

\bibitem{pallisterlinear}
Joe Pallister.
\newblock Linear relations and integrability for cluster algebras from affine
  quivers.
\newblock {\em Glasgow Mathematical Journal}, page 1–38, 2020.

\bibitem{zamolodchikov}
Al~B Zamolodchikov.
\newblock On the thermodynamic {Bethe} ansatz equations for reflectionless
  {ADE} scattering theories.
\newblock {\em Physics Letters B}, 253(3-4):391--394, 1991.

\end{thebibliography}
\end{document}